\documentclass[a4paper,notitlepage,twocolumn,superscriptaddress,longbibliography,aps,nofootinbib,prb]{revtex4-2}

\usepackage[utf8]{inputenc}
\usepackage{braket}
\usepackage{amsmath,amsthm,amssymb,amsfonts}
\usepackage{mathtools}
\usepackage{graphicx}
\usepackage[breaklinks=true]{hyperref}
\usepackage{comment}
\usepackage{color}
\usepackage{makecell}
\usepackage{enumitem}

\usepackage[dvipsnames]{xcolor}
\usepackage{braket}
\usepackage{bbm}
\usepackage{bm}
\usepackage{longtable}
\usepackage[bb=boondox]{mathalfa}
\usepackage{adjustbox}
\usepackage{array}
\usepackage{multirow}
\usepackage{tabularx}
\usepackage{float}

\usepackage{array}
\newcolumntype{C}[1]{>{\centering\let\newline\\\arraybackslash\hspace{0pt}}m{#1}}

\usepackage{tikz}
\usetikzlibrary{shapes,arrows,patterns}
\usepackage{tikz-cd}

\newcommand{\tikzmark}[1]{\tikz[overlay,remember picture] \node[xshift=3mm] (#1) {};}
\newcommand{\tikzmarks}[1]{\tikz[overlay,remember picture] \node[xshift=-.5mm] (#1) {};}

\tikzset{
    triangle/.style={
        draw,
        shape border rotate=0,
        regular polygon,
        regular polygon sides=3,
        node distance=1cm,
        minimum height=.8cm
    }
}
\tikzset{
    square/.style={
        draw,
        shape border rotate=0,
        regular polygon,
        regular polygon sides=4,
        node distance=1cm,
        minimum height=.8cm
    }
}

\tikzset{
    striangle/.style={
        draw,
        shape border rotate=0,
        regular polygon,
        regular polygon sides=3,
        node distance=.5cm,
        minimum height=.1cm
    }
}
\tikzset{
    ssquare/.style={
        draw,
        shape border rotate=0,
        regular polygon,
        regular polygon sides=4,
        node distance=.5cm,
        minimum height=.2cm
    }
}

\definecolor{gray}{gray}{0.5}
\definecolor{G}{rgb}{0.6,0,0}
\definecolor{J}{rgb}{0.6,0,0.6}
\definecolor{O}{rgb}{0,0,0.6}

\newcommand{\X}{Z}
\newcommand{\I}{a}
\newcommand{\J}{b}
\newcommand{\Ug}[1]{\mathcal{U}(#1)}
\newcommand{\Ugd}[1]{\mathcal{U}^\dag(#1)}
\newcommand{\Ugm}[1]{\mathcal{U}^{-1}(#1)}
\newcommand{\Ung}[1]{\mathcal{V}(#1)}
\newcommand{\Ungd}[1]{\mathcal{V}^\dag(#1)}
\newcommand{\M}{g}
\newcommand{\W}{M}
\newcommand{\al}{\eta}
\newcommand{\st}{\;:\;}
\newcommand{\G}{G}

\newtheorem{example}{Example}

\newcommand{\ii}{\mathrm{i}}
\newcommand{\ie}{{\it i.e.},\ }
\newcommand{\eg}{{\it e.g.},\ }

\newcommand{\id}{\mathbb{1}} %identity

\usetikzlibrary{decorations.markings,arrows,decorations.pathreplacing,calc,intersections,through,backgrounds}

\tikzset{->-/.style={decoration={markings,mark=at position #1 with {\arrow{>}}},postaction={decorate}}}
\tikzset{
	partial ellipse/.style args={#1:#2:#3}{
		insert path={+ (#1:#3) arc (#1:#2:#3)}
	}
}

\begin{document}
\title{Generalization of group-theoretic coherent states for variational calculations}

\author{Tommaso Guaita}
\email{tommaso.guaita@mpq.mpg.de}
\affiliation{Max-Planck-Institut für Quantenoptik, Hans-Kopfermann-Str.~1, 85748 Garching, Germany}
\affiliation{Munich Center for Quantum Science and Technology, Schellingstr.~4, 80799 München, Germany}

\author{Lucas Hackl}
\email{lucas.hackl@unimelb.edu.au}
\affiliation{School of Mathematics and Statistics \& School of Physics, The University of Melbourne, Parkville, VIC 3010, Australia}
\affiliation{QMATH, Department of Mathematical Sciences, University of Copenhagen, Universitetsparken 5, 2100 Copenhagen, Denmark}

\author{Tao Shi}
\affiliation{CAS Key Laboratory of Theoretical Physics, Institute of Theoretical Physics, Chinese Academy of Sciences, Beijing 100190, China}
\affiliation{CAS Center for Excellence in Topological Quantum Computation, University of Chinese Academy of Sciences, Beijing 100049, China}

\author{Eugene Demler}
\affiliation{Lyman Laboratory, Department of Physics, Harvard University, 17 Oxford St., Cambridge, MA 02138, USA}

\author{J. Ignacio Cirac}
\affiliation{Max-Planck-Institut für Quantenoptik, Hans-Kopfermann-Str.~1, 85748 Garching, Germany}
\affiliation{Munich Center for Quantum Science and Technology, Schellingstr.~4, 80799 München, Germany}

\begin{abstract}
We introduce new families of pure quantum states that are constructed on top of the well-known Gilmore-Perelomov group-theoretic coherent states. We do this by constructing unitaries as the exponential of operators quadratic in Cartan subalgebra elements and by applying these unitaries to regular group-theoretic coherent states. This enables us to generate entanglement not found in the coherent states themselves, while retaining many of their desirable properties. Most importantly, we explain how the expectation values of physical observables can be evaluated efficiently. Examples include generalized spin-coherent states and generalized Gaussian states, but our construction can be applied to any Lie group represented on the Hilbert space of a quantum system. We comment on their applicability as variational families in condensed matter physics and quantum information.
\end{abstract}

\maketitle

\section{Introduction}
Families of many-body quantum states play an important role in many contexts of quantum science. They are studied in quantum information because they have interesting entanglement structures or because they can be shown to be useful for specific computational tasks. In quantum many-body physics they underlie many collective phenomena and are particularly important for variational methods, both in classical and in quantum computations. For all these applications, the states of these families should be either easy to prepare experimentally (\eg in a quantum computer) or it should be easy to calculate with them classically. Several families that fulfil one or both properties have been studied. For example, product states, Gaussian states and matrix product states (MPS) fulfil both criteria. However, they have limited potential to accomplish the tasks above. For instance, product states do not have correlations at all and Gaussian states have them only in limited forms, while MPS are specifically constructed for 1D geometries.

The goal of this manuscript is to extend some existing families, such that they continue to satisfy both properties above, but contain more correlations or can be used for higher dimensional systems. To do this we base ourselves on two observations: (i) there exist classes of states that extend Gaussian states~\cite{shi2018variational} or spin product states~\cite{kitagawa_squeezed_1993,Foss-Feig_2013_Noneq} to contain more correlations while continuing to admit easy computations of expectation values; (ii) Gaussian states, bosonic coherent states and some classes of product states can all be understood within a unified framework based on Lie group theory.

\begin{figure}
	\begin{center}
	\begin{tikzpicture}
	\draw[-,thick] (-1.0,0) -- (1.0,0);
	\draw[-,thick] (0.35,2.33826) -- (1.35,0.6062);
	\draw[-,thick] (-0.35,2.33826) -- (-1.35,0.6062);

		\node[draw,align=center,left,G] at (-1.2,0) {\color{black}\textbf{ Bosons:}\\ $\hat{H}^{(\mathrm{b})}_l=\ii(\hat{a}_l^\dag \hat{a}_l +\frac{1}{2})$};
		
		\node[draw,align=center,right,O] at (1.2,0) {\color{black}\textbf{Fermions:}\\ $\hat{H}^{(\mathrm{f})}_n=\ii(\hat{c}_n^\dag \hat{c}_n -\frac{1}{2})$};
		
		\node[draw,align=center,J] at (0,2.9444) {\color{black}\textbf{Spins:}\\ $\hat{H}^{(\mathrm{s})}_k=\frac{\ii}{2} \hat{\sigma}^k_3$};
		
		\draw[] (0,-1) node[align=center]{$\exp(-\frac{\ii}{2}\W^{ln}{\color{G}\hat{H}^{(\mathrm{b})}_l}{\color{O}\hat{H}^{(\mathrm{f})}_n})$};
		
		\draw[] (2.7,1.5) node[align=center]{$\exp(-\frac{\ii}{2}\W^{kn}{\color{J}\hat{H}^{(\mathrm{s})}_k}{\color{O}\hat{H}^{(\mathrm{f})}_n})$};
		
		\draw[] (-2.7,1.5) node[align=center]{$\exp(-\frac{\ii}{2}\W^{lk}{\color{G}\hat{H}^{(\mathrm{b})}_l}{\color{J}\hat{H}^{(\mathrm{s})}_k})$};
		
		\draw[] (0,1) node[align=center]{$\Ung{M}$};
	
	\end{tikzpicture}
	\end{center}
	\caption{We show schematically how a unitary operator $\Ung{M}$ can generate entanglement in composite systems between different sectors, \eg bosonic, fermionic or spin sectors. %$\Ung{M}$ will contain products of Cartan sub-algebra operators coming either from the spin and boson sectors, from the spin and fermion sectors or from the fermion and boson sectors.
	}
	\label{fig:triangle}
\end{figure}
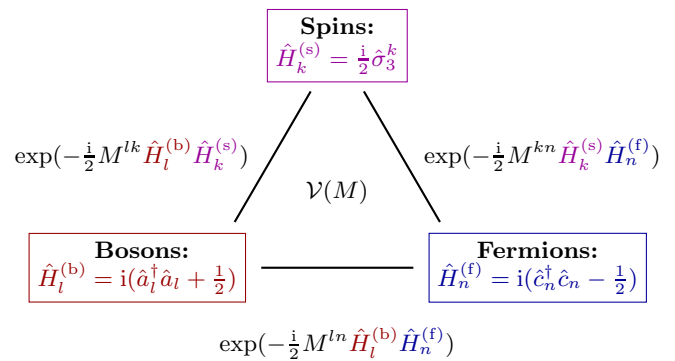

This unification was understood independently by Gilmore~\cite{gilmore1972geometry,gilmore1974properties} and Perelomov~\cite{perelomov1972coherent,perelomov2012generalized}, leading to the definition of so-called \emph{group-theoretic coherent states}. These are defined by the action of a unitary representation of a Lie group on a fixed reference state. The properties of the ensuing family of states are fully encoded in the algebraic properties of the chosen group and representation. Several frequently used families of quantum states can be understood as instances of group-theoretic coherent states resulting from different choices of Lie groups. Standard bosonic coherent states arise from the group of translations, bosonic and fermionic Gaussian states arise from representations of the groups $\mathrm{Sp}(2N,\mathbb{R})$ and $\mathrm{O}(2N,\mathbb{R})$, while atomic coherent states~\cite{arecchi1972atomic} arise from the two dimensional representation of $\mathrm{SU}(2)$.

Exploiting these available group-theoretical structures, we thus consistently extend all families of group-theoretic coherent states to include states that go beyond the coherent state paradigm, while still maintaining the property of efficient computation of expectation values. We achieve this by applying to them a single unitary transformation $\Ung{\W}=\exp(-\frac{\ii}{2}\W^{\I\J}\hat{H}_\I\hat{H}_\J)$, where $\hat{H}_\I$ represents a so-called Cartan subalgebra operator and the matrix $M$ contains additional variational parameters. This construction is inspired by the extensions of Gaussian states defined in~\cite{shi2018variational}.

As said, the specific form of this extension is designed to preserve the desirable feature of being able to compute expectation values efficiently. In fact, all necessary operations are performed in terms of objects (matrices and vectors) whose dimension is at most the one of the Lie group. In most examples, this dimension scales polynomially with the size of the considered system, making our methods feasible even for studying large systems and exploring the thermodynamic limit.
While satisfying this constraint, the extension also enlarges the range of available types of quantum correlations, going thus beyond mean field treatments, such as the Landau-Lifshitz equations~\cite{lakshmanan2011fascinating}. Indeed, the exponent of $\Ung{M}$, which is quadratic in algebra operators, can represent structures not present in coherent states. For example, it can be used to introduce non trivial density-density correlations in Gaussian states or spin-spin correlations in spin systems. Furthermore, in composite systems it can produce entanglement between different types of degrees of freedom (spins, bosons, fermions) as it can contain products of Cartan subalgebra operators from the different sectors, as sketched in figure~\ref{fig:triangle}. 

The proposed construction is very general, in the sense that it can be applied to group-theoretic coherent states associated to any choice of Lie group. For this reason, we will give all definitions in a sufficiently general language that does not refer to a specific Lie group and algebra. To make the rather formal construction more concrete, we will illustrate each step for two paradigmatic examples, namely spin-$\frac{1}{2}$ coherent states and bosonic Gaussian states.\\

This manuscript is structured as follows: In section~\ref{sec:group-theoretic-coherent-states-global}, we review the construction of group-theoretic coherent states according to the insight of Gilmore and Perelomov. In section~\ref{sec:generalized-family-global}, we define our generalization of group-theoretic coherent states and show how any expectation value with respect to those states can be brought into a certain standard form. In section~\ref{sec:efficient-computation-global}, we then explain how expectation values in the previously introduced standard form can be evaluated efficiently. In section~\ref{sec:summary}, we summarize our findings and give an outlook of where we believe they will be most useful. In appendix~\ref{app:ex-su2} and~\ref{app:ex-gaussian}, we provide a detailed discussions of the examples mentioned in the main text, namely spin-$\frac{1}{2}$ coherent states and bosonic Gaussian states. For completeness, in Appendix~\ref{app:ex-fermionic-gaussian} we also give more details about the case of fermionic Gaussian states, another paradigmatic example to which our construction can be applied.

\section{Group-theoretic coherent states}
\label{sec:group-theoretic-coherent-states-global}
In this section, we review the basic definition and properties of group-theoretic coherent states based on~\cite{zhang1990coherent} and following the conventions of~\cite{hackl2020geometry}, where we studied their geometric properties.

We consider a semi-simple Lie group $\mathcal{G}$ with Lie algebra $\mathfrak{g}$. Let $\mathcal{U}$ be a unitary representation of $\mathcal{G}$ on the Hilbert space $\mathcal{H}$, \ie $\mathcal{U}(\M)$ is a unitary operator on $\mathcal{H}$ for every group element $\M\in\mathcal{G}$, such that \begin{equation}
    \Ug{\M_1}\Ug{\M_2} = \Ug{\M_1 \M_2} \quad \forall \M_1,\M_2\in \mathcal{G}\,.
    \label{eq:group-multiplication}
\end{equation}

The representation of the group induces a corresponding representation of the algebra. Indeed, for group elements $\M$ sufficiently close to the identity, it is possible to write $\Ug{\M}=\mathrm{exp}(K^i \hat{\X}_i)$, where $\hat{\X}_i$ is a set of anti-Hermitian operators representing a basis of the algebra $\mathfrak{g}$ and $K^i$ are real coefficients. We have the commutation relations\footnote{Note that here, as in the rest of the paper, we use Einstein's convention of summing implicitly over all repeated indices.} 
\begin{equation}
    [\hat{\X}_i,\hat{\X}_j]=c_{ij}^k \hat{\X}_k\,,
    \label{eq:structure-constants}
\end{equation}
fixed by the structure constants $c_{ij}^k$ of the algebra. The action of $\Ug{\M}$ on the operators $\hat{\X}_i$ follows the adjoint representation of the group. More precisely, we have
\begin{equation}
    \Ugm{\M}\, \hat{\X}_i \,\Ug{\M} = \mathrm{Ad}(\M)_i^j\, \hat{\X}_j\,,
    \label{eq:adjoint-representation}
\end{equation}
\ie $\Ugm{\M}\, \hat{\X}_i \,\Ug{\M}$ is just a linear combination of operators $\hat{\X}_i$ with the coefficients given by the adjoint matrix $\mathrm{Ad}(\M)_i^j$, which is a fixed property of the group\footnote{In particular, if we can write $\Ug{\M}=\mathrm{exp}(K^i \hat{\X}_i)$, then it is straightforward to see that $\mathrm{Ad}(\M)_i^j= \left[\exp \mathrm{ad}(K)\right]_i^j$ where the matrix $\mathrm{ad}(K)$ is given by $\mathrm{ad}(K)_i^j=K^k c_{ki}^j$. For a more complete discussion see Appendix~\ref{app:adjoint-rep}.\label{foot:adjoint-rep}}.

The set $\mathcal{M}_\phi$ of group-theoretic coherent states is then defined as the set of states obtained by acting with all possible $\Ug{\M}$ on a fixed reference state $\ket{\phi}\in\mathcal{H}$, \ie
\begin{equation}
    \mathcal{M}_\phi=\{\;\Ug{\M}\ket{\phi}\st \M\in\mathcal{G}\}\subset\mathcal{H}\,.
    \label{eq:definition-group-theoretic-coherent-states}
\end{equation}

$\mathcal{M}_\phi$ is determined by the choice of the group $\mathcal{G}$, of its representation $\mathcal{U}$ and of the reference state $\ket{\phi}$. The elements of $\mathcal{M}_\phi$ are parametrized by group elements $\M$. This parametrization may entail some redundancies, as there might exist in $\mathcal{G}$ a stabilizer subgroup for $\ket{\phi}$
\begin{equation}
    S_\phi = \{\;\M \st \Ug{\M}\ket{\phi}=e^{\ii\theta} \ket{\phi} \}\,,
\end{equation}
\ie a set of group transformations that leave $\ket{\phi}$ unchanged up to an overall phase, which is irrelevant for what concerns the definition of quantum states. The set of inequivalent group-theoretic coherent states is then isomorphic to the quotient $G/S_\phi$.

For our purposes, it is necessary to restrict the possible choices for the reference state $\ket{\phi}$. We will indeed assume that $\ket{\phi}$ is a so-called lowest weight state of the representation $\mathcal{U}$. To understand what is meant by this it is necessary to give some more details about the structure of the algebra operators~\cite{georgi2018lie,knapp1996liegroups}. We will explain this in the rest of this section.

It is always possible to pick a set of $\ell$ linearly independent mutually commuting anti-Hermitian operators $\hat{H}_\I=H_\I^i\hat{\X}_i$,  defined by $H_\I^i\in\mathbb{R}$ for $\I=1,\dots, \ell$, such that $[\hat{H}_\I,\hat{H}_\J]=0$. In the standard theory of Lie algebras, the space spanned by real linear combinations of $\hat{H}_\I$, which we will indicate with $\mathfrak{h}$, is known as a \emph{Cartan subalgebra} of $\mathfrak{g}$. The choice of $\mathfrak{h}$ is not unique, however all possible choices are isomorphic and will therefore have the same dimension $\ell$, known as the \emph{rank} of the algebra. A given a choice of Cartan subalgebra identifies the following structures:
\begin{itemize}
    \item There exist real vectors $\al=(\al_1,\dots,\al_\ell)\in\mathbb{R}^\ell$ and corresponding operators $\hat{E}_\al$ such that
    \begin{equation}
        [\hat{H}_\I,\hat{E}_{\al}]=\ii \al_\I \hat{E}_{\al}\,.
        \label{eq:adjoint-action-H-on-E}
    \end{equation}
    The operators $\hat{E}_\al$ will be linear combinations of $\hat{\X}_i$, however they will in general be complex linear combinations and therefore will not be anti-Hermitian operators.
    \item The vectors $\al$ are known as \emph{roots} of the algebra and the operators $\hat{E}_\al$ as \emph{root space operators}. There is a finite set of non-zero roots which we indicate as $\Delta$. The roots always come in pairs $(\al,-\al)$. One can choose a conventional ordering of the roots such that they split into the two disjoint sets of positive roots $\Delta_+$ and negative roots $\Delta_-$, with $\Delta=\Delta_+ \cup \Delta_-$ and $-\al\in\Delta_-$ for every $\al\in\Delta_+$.
    \item Let us indicate with $\mathfrak{g}^\mathbb{C}$ the space of all complex linear combinations of algebra elements $\hat{\X}_i$, which is known as the \emph{complexified} Lie algebra. The operators $\hat{H}_\I$ together with the operators $\hat{E}_\al$ span $\mathfrak{g}^\mathbb{C}$ under complex linear combinations.
\end{itemize}

A Hilbert space vector $\ket{\mu}\in\mathcal{H}$ is called a \emph{weight} vector of the representation if it is a common eigenstate of all Cartan subalgebra operators $\hat{H}_\I$, \ie $\hat{H}_\I\ket{\mu}=\ii \mu_\I\ket{\mu}$ for some number $\mu_\I\in\mathbb{R}$ $\forall \I$. Among the weight vectors $\ket{\mu}$ there is a unique one, called the \emph{lowest} weight vector, such that $\hat{E}_\al \ket{\mu}=0$ for all negative roots $\al\in\Delta_-$. From now on we assume that the reference state $\ket{\phi}$ that appears in the definition~\eqref{eq:definition-group-theoretic-coherent-states} of group-theoretic coherent states is a lowest weight vector $\ket{\mu}$ for a given choice of Cartan subalgebra and root ordering.

\begin{example}[Spin-$\frac{1}{2}$ coherent states]
\label{ex:spin-coherent-states}
Spin-$\frac{1}{2}$ coherent states are defined with respect to the group $\mathrm{SU}(2)$ and algebra $\mathfrak{su}(2)$, represented as complex $2$-by-$2$ matrices. For the algebra, we choose the basis $Z_i=\ii \sigma_i$ with $\sigma_i$ being the well-known Pauli matrices. The rank of $\mathfrak{su}(2)$ is $1$ and, as conventional, we choose $\hat{H}=\frac{\ii}{2}\sigma_3$ as basis of the Cartan subalgebra $\mathfrak{h}$. For this choice, we have the roots $\pm\eta=\pm1$, with the respective root space operators
\begin{align}
    \hat{E}_{\pm\al}&=\hat{\sigma}_\pm=\frac{1}{2\sqrt{2}}(\hat{\sigma}_1\pm\ii\hat{\sigma}_2)=\frac{1}{2\sqrt{2}}(-\ii\hat{\X}_1\pm\hat{\X}_2)\,.
\end{align}
The resulting weight vectors are $\ket{\uparrow}$ and $\ket{\downarrow}$ because they are the eigenvectors of $\hat{H}=\frac{\ii}{2}\sigma_3$. Due to $\hat{E}_{-\al}\ket{\downarrow}=0$, the state $\ket{\downarrow}$ is the lowest weight vector, which we thus choose as reference state. The family of group-theoretic coherent states results then from applying all possible group elements $\mathcal{U}\in\mathrm{SU}(2)$ and is given by
\begin{equation}
    \mathcal{M}_{\mathrm{SU}(2)}=\{ e^{\ii K^i \hat{\sigma}_i}\ket{\downarrow} \st K\in\mathbb{R}^3\}\,.
\end{equation}
This construction can be readily extended to a system of $N$ spin-$\frac{1}{2}$, in which case the Cartan algebra will be composed of $N$ operators $\hat{H}_k=\frac{\ii}{2} \hat{\sigma}^k_3$, one for each spin $k$, and the lowest weight vector will be $\ket{\mu}=\ket{\downarrow\dots\downarrow}$.
\end{example}

\begin{example}[Bosonic Gaussian states]
\label{ex:gaussian-states}
The well-known Gaussian states for a system of $N$ bosonic modes can be understood as the group-theoretic coherent states arising from the algebra of all anti-Hermitian operators $\hat{Q}$ that are quadratic in the canonical creation and annihilation operators $\hat{a}^\dag_k$ and $\hat{a}_k$. The corresponding unitary group is the one of all operators that can be written as $\mathcal{U}=e^{\hat{Q}}$.\\
Within the algebra of quadratic operators we can choose the Cartan operators
\begin{equation}
    \hat{H}_k=\ii(\hat{a}_k^\dag \hat{a}_k +\frac{1}{2})\,,
\end{equation}
and root space operators
\begin{subequations}
\begin{align}
    \hat{E}_{+\al^{(k,l)}}&=\ii\hat{a}^\dag_k\hat{a}^\dag_l,&\hat{E}_{-\al^{(k,l)}}&=\ii\hat{a}_k\hat{a}_l,& k\leq l\\
    \hat{E}_{+\tilde{\al}^{(k,l)}}&=\hat{a}^\dag_k\hat{a}_l,&\hat{E}_{-\tilde{\al}^{(k,l)}}&=\hat{a}_k\hat{a}^\dag_l,&k<l
\end{align}
\end{subequations}
corresponding to the root vectors $\al^{\!(k,l)}_{\,\I}=(\delta_{ak}+\delta_{al})$ and $\tilde{\al}^{(k,l)}_{\,\I}=(\delta_{\I k}-\delta_{\I l})$.
The lowest weight vector of this representation is the Fock vacuum $\ket{0}$ as it is an eigenstate of all $\hat{H}_k$ and is annihilated by all $\hat{E}_{-}$. The corresponding group-theoretic coherent states are then all states that can be written as $e^{\hat{Q}}\ket{0}$, which we recognise as conventional bosonic Gaussian states.
The algebra of quadratic operators $\hat{Q}$ and the corresponding group of unitaries $\mathcal{U}$ can be recognised as infinite dimensional representations of the Lie algebra 
$\mathfrak{sp}(2N,\mathbb{R})$ and Lie group $\mathrm{Sp}(2N,\mathbb{R})$. For more details on this and on how to parametrise the unitaries $\Ug{S}$ and algebra operators $\hat{Q}(K)$ in terms of matrices $S\in\mathrm{Sp}(2N,\mathbb{R})$ and $K\in\mathfrak{sp}(2N,\mathbb{R})$ see Appendix~\ref{app:ex-gaussian}.
\end{example}

\section{Generalized group-theoretic coherent states and standard form of expectation values}
\label{sec:generalized-family-global}
In this section, we will first define new families of states, which we refer to as \emph{generalized group-theoretic coherent states}, that extend the families of group-theoretic coherent states described in the previous section. In the second part, we will then show how the expectation value of arbitrary operators (written as power series of reference operators) can be brought into a \emph{standard form}, which can then be evaluated efficiently.%The aim is to define states that go beyond coherent states in a non-trivial way, possessing more types of quantum correlations, but for which computing the relevant physical quantities remains a sufficiently simple task.

\subsection{Definition}
\label{sec:generalized-family}
We choose a Cartan subalgebra $\mathfrak{h}\subset\mathfrak{g}$, spanned by the operators $\hat{H}_\I$ as defined in the previous section. Let us then consider the unitary operator
\begin{equation}
    \Ung{\W}=\exp\left(\frac{\ii}{2}\W^{\I\J}\hat{H}_\I\hat{H}_\J\right)\,.
    \label{eq:definition-U_W}
\end{equation}
The real symmetric matrix $\W^{\I\J}$ defines a bilinear form on $\mathfrak{h}$ and contains $\ell(\ell+1)/2$ real parameters that define the operator.

The exponent of~\eqref{eq:definition-U_W} is not an element of the Lie algebra $\mathfrak{g}$, as it is quadratic in the basis operators $\hat{\X}_i$. Consequently, $\Ung{\W}$ is not a group transformation and the product of more operators of this type does not follow a group multiplication rule. Furthermore, the action of a transformation $\Ung{\W}$ will in general take an element of $\mathcal{M}_\phi$ out of the set of group-theoretic coherent states.

We now define the class of generalized group-theoretic coherent states as the set of states of the form
\begin{align}
    \ket{\psi(\M_1,\M_2,\W)}=\Ug{\M_1}\,\Ung{\W}\,\Ug{\M_2}\ket{\mu}\,.
    \label{eq:definition-generalized-coherent-states}
\end{align}
The states are conveniently parametrized by two group elements $\M_1$ and $\M_2$ and one bilinear form $\W$, although this parametrization will contain several redundancies. Similarly to group-theoretic coherent states, this class of states is determined by the choice of the group $\mathcal{G}$ and of its representation $\mathcal{U}$ on Hilbert space. In the case of compact Lie groups any choice of Cartan subalgebra and lowest weight state $\ket{\mu}$ will define the same family of states\footnote{This is because in this case all Cartan subalgebras and lowest weight states are equivalent up to group unitary transformations, which can be absorbed in to the parameters $g_1$ and $g_2$. In the case of non-compact Lie groups there may instead exist unitarily inequivalent classes of Cartan subalgebras. Their choice is therefore relevant. Note that the choice with respect to which operator~\eqref{eq:definition-U_W} is defined may even be different from the one with respect to which the lowest weight state $\ket{\mu}$ is defined.}.

\begin{example}[Generalized spin-$\frac{1}{2}$ coherent states]
\label{ex:generalized-spin-coherent-states}
Based on Example~\ref{ex:spin-coherent-states}, we consider a system of $N$ spin-$\frac{1}{2}$ degrees of freedom with Cartan algebra spanned by $\hat{H}_k=\frac{\ii}{2}\sigma_3^k$. The unitary operator~\eqref{eq:definition-U_W} takes the form 
\begin{equation}
    \Ung{M}=\exp \left(-\frac{\ii}{8}M_{kl}\,\hat{\sigma}_3^k \hat{\sigma}^l_3\right)\,,
\end{equation}
for any given $N\times N$ real symmetric matrix $M$. The \emph{generalized} spin-$\frac{1}{2}$ coherent states take the form
\begin{align}
    \ket{\psi(K_1,K_2,M)}&=\Ug{K_1}\,\Ung{M}\,\Ug{K_2}\ket{\downarrow\cdots\downarrow}\,,
\end{align}
where, similarly to Example~\ref{ex:spin-coherent-states} and as explained in more detail in Appendix~\ref{app:ex-su2}, the group unitaries are defined as
\begin{equation}
    \Ug{K}=\exp\left(\ii K^{i,k} \hat{\sigma}^k_i\right)\,,
\end{equation}
with the coefficients $K^{i,k}$ taking values for $i=1,2,3$ and for each spin $k=1,\dots,N$.
\end{example}

\begin{example}[Generalized bosonic Gaussian states]
\label{ex:generalized-gaussian-states}
Based on Example~\ref{ex:gaussian-states}, we consider a system of $N$ bosonic modes with Cartan algebra spanned by $\hat{H}_k=\ii(\hat{a}_k^\dag \hat{a}_k +\frac{1}{2})$.
The unitary operator~\eqref{eq:definition-U_W} takes the form 
\begin{equation}
    \Ung{M}=\exp \left(-\frac{i}{2} M^{kl} (\hat{a}_k^\dag \hat{a}_k +\tfrac{1}{2})(\hat{a}_l^\dag \hat{a}_l +\tfrac{1}{2}) \right)\,,
\end{equation}
for any given $N\times N$ real symmetric matrix $M$. The \emph{generalized} bosonic Gaussian states take the form
\begin{align}
    \ket{\psi(S_1,S_2,M)}&=\Ug{S_1}\,\Ung{M}\,\Ug{S_2}\ket{0}\,,
    \label{eq:generalized-gaussian-states-ex}
\end{align}
where $\Ug{S}$ are the Gaussian unitaries discussed in Example~\ref{ex:gaussian-states} and defined more precisely in Appendix~\ref{app:ex-gaussian}. We recognize that these states constitute one of the classes of non-Gaussian states previously introduced in~\cite{shi2018variational}.
\end{example}

\subsection{Entangling degrees of freedom in composite systems}

The construction of group-theoretic coherent states is possible also in the case in which different groups act on different sectors of a composite system. In this case the construction of \emph{generalized} group-theoretic coherent states is particularly useful, because, as mentioned in the introduction,  
it enables us to entangle and correlate the different types of degrees of freedom in the system, such as spins, bosons and fermions. This provides a distinct advantage over coherent states alone, which are always product states over the different system components, described by the different groups (special unitary group for spin, symplectic group for bosons, orthogonal group for fermions).

More precisely, let us assume that we have two semi-simple Lie groups $\mathcal{G}_1$ and $\mathcal{G}_2$, such that the respective representations act on a tensor product of Hilbert spaces $\mathcal{H}=\mathcal{H}_1\otimes\mathcal{H}_2$ and thus commute with each other, \ie we have a representation of the product group $\mathcal{G}=\mathcal{G}_1\times\mathcal{G}_2$ with Lie algebra $\mathfrak{g}=\mathfrak{g}_1\oplus \mathfrak{g}_2$. By applying the construction of group-theoretic coherent states, we will find that the Cartan subalgebra $\mathfrak{h}=\mathfrak{h}_1\oplus \mathfrak{h}_2$ is the direct sum of the respective Cartan subalgebras. Following our definition of generalized coherent states, the transformation $\Ung{\W}$ will contain three terms, \ie
\begin{align}
\begin{split}
    \frac{\ii}{2}M^{ab}\hat{H}_a\hat{H}_b&=\frac{\ii}{2}\big(M^{ab}_{(1)}\hat{H}^{(1)}_a\hat{H}^{(1)}_b+M^{ab}_{(2)}\hat{H}^{(2)}_a\hat{H}^{(2)}_b\\
    &\quad+2M^{ab}_{(12)}\hat{H}^{(1)}_a\hat{H}^{(2)}_b\big)\,,\label{eq:V-composite}
\end{split}
\end{align}
where $H_a^{(i)}\in\mathfrak{h}_i$. We thus see explicitly that the last term is a product of Cartan generators associated to the two different original groups. As our representation acts on a tensor product, this last term in $\Ung{\W}$ will be responsible for entangling degrees of freedom associated to different parts of a composite system. This is particularly relevant when $\mathcal{G}_1$ and $\mathcal{G}_2$ are associated to different types of physical degrees of freedom, such as spins, bosons and fermions.

\begin{example}[Entangling spin-$\frac{1}{2}$ and bosonic systems]
Let us consider a system composed of $N$ spin-$\frac{1}{2}$ degrees of freedom, as described in Example~\ref{ex:spin-coherent-states}, and $\tilde{N}$ bosonic modes, as described in Example~\ref{ex:gaussian-states}. The total Lie group acting on it will be given by $\mathcal{G}=\mathrm{SU}(2)^N\times\mathrm{Sp}(2\tilde{N},\mathbb{R})$. The corresponding Cartan subalgebra is given by the span of all the operators
\begin{equation}
    \hat{H}^{(1)}_k=\frac{\ii}{2} \hat{\sigma}^k_3,\hspace{40pt} \hat{H}^{(2)}_k=\ii(\hat{a}_k^\dag \hat{a}_k +\frac{1}{2})\,.
\end{equation}
Consequently the unitary $\Ung{M}$ takes the form
\begin{align}
\begin{split}
    \Ung{M}&=\exp\left[-\frac{\ii}{8}M^{(1)}_{kl}\,\hat{\sigma}_3^k \hat{\sigma}^l_3 \right. \\
    &\hspace{50pt}-\frac{i}{2} M^{kl}_{(2)} (\hat{a}_k^\dag \hat{a}_k +\tfrac{1}{2})(\hat{a}_l^\dag \hat{a}_l +\tfrac{1}{2}) \\
    &\hspace{80pt}\left.-\frac{i}{2}M_{kl}^{(12)}\,\hat{\sigma}_3^k (\hat{a}_l^\dag \hat{a}_l +\tfrac{1}{2}) \right] \,.
\end{split}
\end{align}
In particular we see that the last term generates entanglement between the spin and bosonic degrees of freedom.
\end{example}

\subsection{Standard form of expectation values}
\label{sec:standard-form}
Our definition was carefully chosen, such that we can efficiently compute the expectation value of physical observables $\hat{\mathcal{O}}$ of interest (\eg Hamiltonians). Here, we assume that the group was chosen, such that $\hat{\mathcal{O}}$ can be expressed as a polynomial in the operators $\hat{\X}_i$, which can be accomplished in most physical systems. Then, any such expectation value can be brought into the standard form
\begin{equation}
    \braket{\psi|\hat{\mathcal{O}}|\psi}=\sum_{n,\{i\}}C^{i_1\dots i_n}\braket{\mu|\,\Ug{\M_n}\,\hat{\X}_{i_1}\cdots\hat{\X}_{i_n}|\mu} \,.
    \label{eq:expectation-U-Xi}
\end{equation}
To reach this standard form, we need to commute $\mathcal{U}_1\equiv\Ug{g_1}$, $\mathcal{V}$ and $\mathcal{U}_2\equiv\Ug{g_2}$ through the operators $\hat{Z}_{i}$ that appear in $\hat{\mathcal{O}}$ according to
\begin{equation}
    \braket{\psi|Z_{i_1}\dots Z_{i_n}|\psi}=\langle\mu|\mathcal{U}_2^\dagger\mathcal{V}^\dagger\mathcal{U}_1^\dagger\,\tikzmarks{u1t}\tikzmarks{vt}\tikzmarks{u2t}\, Z_{i_1}\dots Z_{i_n}\tikzmark{u1}\mathcal{U}_1\tikzmark{v}\mathcal{V}\tikzmark{u2}\mathcal{U}_2|\mu\rangle\,.
    \begin{tikzpicture}[overlay,remember picture,out=315,in=225,distance=0.5cm]
    \draw[<-,shorten >=6pt,shorten <=3pt] (u1t.center) to (u1.center);
    \draw[<-,shorten >=6pt,shorten <=3pt] (vt.center) to (v.center);
    \draw[<-,shorten >=6pt,shorten <=3pt] (u2t.center) to (u2.center);
  \end{tikzpicture}
  \label{eq:commutation-with-arrows}
\end{equation}
This will only transform the operators $\hat{Z}_i$ or generate additional group unitaries $\mathcal{U}(g_i)$, which can all be collected to the left to form the single unitary $\mathcal{U}(g_n)$. To do this, we need the following two commutation rules:
\begin{itemize}
    \item \textbf{Commuting $\mathcal{U}$ with $\X_i$:}\\
    From~\eqref{eq:adjoint-representation} we have that commuting group transformations with algebra operators only gives rise to linear combinations according to
    \begin{align}
        \hat{\X}_i \, \Ug{\M}=\mathrm{Ad}(\M)_i^j \,  \Ug{\M}\,\hat{\X}_j\,.\label{eq:commute-left}
    \end{align}
    \item \textbf{Commuting $\mathcal{V}$ with $\X_i$:}\\
    Even though $\Ung{\W}$ is not a group transformation, its action on algebra elements has a simple form. Indeed, from relation~\eqref{eq:adjoint-action-H-on-E} it follows that
    \begin{align}
    \begin{split}
         \hat{E}_\al \:\Ung{\W}&=\Ung{\W}\:e^{\al_\I \W^{\I\J}\hat{H}_\J-\frac{\ii}{2}\al_\I \W^{\I\J}\al_\J}\hat{E}_\al\\
        &=\Ung{\W}\:e^{\ii\theta_{\al}}\,\Ug{e^{K_{\al}}}\hat{E}_\al\,,
    \end{split}\label{eq:action-U_W-on-E}
    \end{align}
    where in the second line we have recognised that the exponential can be decomposed into a complex phase factor $\theta_\al=-\frac{1}{2}\al_\I\W^{\I\J}\al_\J$ and the exponential of a real linear combination of algebra operators $K_{\al}=\al_\I\W^{\I\J}H_\J$. Furthermore we have that
    \begin{equation}
         \hat{H}_\I \Ung{\W} = \Ung{\W}\hat{H}_\I\,,
        \label{eq:action-U_W-on-H}
    \end{equation}
    as $\Ung{\W}$ is a function exclusively of Cartan subalgebra operators and therefore commutes with $\hat{H}_\I$. As all algebra operators $\hat{\X}_i$ can be expressed as complex linear combinations of operators of the types $\hat{H}_\I$ or $\hat{E}_\al$, it follows that the commutation of $\Ung{\W}$ through $\hat{\X}_i$ will be a linear combination of~\eqref{eq:action-U_W-on-H} and~\eqref{eq:action-U_W-on-E}.
\end{itemize}
By combining a series of operations of these kinds, we can always commute the unitaries $\mathcal{U}_1$, $\mathcal{V}$ and $\mathcal{U}_2$ in~\eqref{eq:commutation-with-arrows} through any monomial of operators $\hat{\X}_i$. They will then combine with the corresponding $\mathcal{U}^\dag_1$, $\mathcal{V}^\dag$ and $\mathcal{U}^\dag_2$ coming from the bra vector $\bra{\psi}$ yielding identities and leaving a linear combination of terms of the form $\Ug{\M}\,\hat{\X}_{j_1}\cdots\hat{\X}_{j_n}$. 

More specifically, the unitaries $\Ung{M}$ will give rise to a series of group transformations $e^{\theta_{\al_i}}\Ug{e^{K_{\al_{i}}}}$ according to~\eqref{eq:action-U_W-on-E}. Then one has to commute all $\Ug{e^{K_{\al_{i}}}}$ to the left using using \eqref{eq:commute-left}, which will produce linear combinations of $\Ug{e^{K_{\eta_i}}}\,\hat{\X}_{j_1}\cdots\hat{\X}_{j_n}$. Once all the group transformations are on the left side, they combine to $\Ug{\M}=\Ug{e^{K_{\al_{i_1}}}}\dots\Ug{e^{K_{\al_{i_n}}}}$. Thus, the action of $\Ung{\W}$ on a monomial of algebra operators $\hat{\X}_i$ will give rise to a polynomial of the same order multiplied with a single group transformation $\Ug{\M}$ from the left.

In summary, any expectation value of an observable $\hat{\mathcal{O}}$ can be brought into the standard form~\eqref{eq:expectation-U-Xi}, whose efficient evaluation will be subject of the next section. This enables the application of a wide range of variational methods when using generalized group-theoretic coherent states as an approximation of the true state of the system\footnote{The more experienced reader will know that to apply the full range of known variational methods to a given family of quantum states (\eg as described in~\cite{hackl2020geometry}), it is not always sufficient to be able to compute the expectation values of the Hamiltonian. It is also necessary to compute quantities involving so-called \emph{tangent vectors}. In appendix~\ref{app:variational-methods}, however, we show that for generalized group-theoretic coherent states also these quantities can be simply brought to the standard form~\eqref{eq:expectation-U-Xi}.}. The specific form of definition~\eqref{eq:definition-U_W} -- which at first sight may appear somewhat arbitrary -- was fundamental for achieving this. Indeed, the inclusion in the exponent of~\eqref{eq:definition-U_W} of algebra elements outside of the Cartan subalgebra or of non-quadratic terms would make it impossible to express the transformations~\eqref{eq:commute-left} and~\eqref{eq:action-U_W-on-E} exclusively in terms of algebra and group operators, and thus would prevent the subsequent calculations.

\begin{example}[Commutation rules for generalized spin-$\frac{1}{2}$ coherent states]
The operators $\mathcal{U}(K)$ and $\Ung{M}$, defined in Example~\ref{ex:generalized-spin-coherent-states} satisfy the following relations:
\begin{equation}
    \left(\begin{array}{c} \hat{\sigma}_1^k \\\hat{\sigma}_2^k \\\hat{\sigma}_3^k \end{array}\right) \,\Ug{K} = \Ug{K}\, e^{-2 K^{i,k} \mathbf{L}_i}\, \left(\begin{array}{c} \hat{\sigma}_1^k \\\hat{\sigma}_2^k \\\hat{\sigma}_3^k \end{array}\right)\,,
    \label{eq:su2-adjoint-example}
\end{equation}
corresponding to~\eqref{eq:commute-left}, where we have the $3$-by-$3$ matrices $(\mathbf{L}_i)_{mn}=\epsilon_{imn}$, with $\epsilon_{imn}$ being the totally antisymmetric tensor; and
\begin{align}
    \hat{\sigma}_3^k \,\Ung{M}&=\Ung{M}\,\hat{\sigma}_3^k\,,\\
    \hat{\sigma}_\pm^k \,\Ung{M}&=\Ung{M}\,e^{-\frac{\ii}{2}M_{kk}}e^{\pm\frac{\ii}{2}M_{kl}\hat{\sigma}_3^l}\hat{\sigma}_\pm^k\,,
\end{align}
corresponding to~\eqref{eq:action-U_W-on-H} and~\eqref{eq:action-U_W-on-E}.
\end{example}

\begin{example}[commutation rules for generalized Gaussian states]
The commutation of $\Ug{S}$, discussed in Examples~\ref{ex:gaussian-states} and~\ref{ex:generalized-gaussian-states}, with any creation or annihilation operator can be achieved through
\begin{equation}
    \Ugd{S}\,\hat{\mathbf{x}}\,\Ug{S}=S\hat{\mathbf{x}}\,,
\end{equation}
where $\hat{\mathbf{x}}=(\hat{q}_1,\dots,\hat{q}_N,\hat{p}_1,\dots,\hat{p}_N)^\intercal$ and $\hat{q}_i=(\hat{a}^\dag_i+\hat{a}_i)/\sqrt{2}$ and $\hat{p}_i=\ii(\hat{a}^\dag_i-\hat{a}_i)/\sqrt{2}$ are canonical quadrature operators. The commutation of $\Ung{M}$, discussed in Examples~\ref{ex:generalized-gaussian-states}, with creation or annihilation operators can be achieved through
\begin{align}
\begin{split}
    \Ungd{M}\hat{a}_k\hat{a}_l \Ung{M}&=e^{-\frac{\ii}{2}(M^{kk}+M^{kl}+M^{lk}+M^{ll})}\\
    &\times e^{-\ii(M^{km}+M^{lm})(\hat{a}_m^\dag \hat{a}_m +\frac{1}{2})}\hat{a}_k\hat{a}_l\,,
\end{split}\\
\begin{split}
    \Ungd{M}\hat{a}_k^\dag\hat{a}_l \Ung{M}&=e^{-\frac{\ii}{2}(M^{kk}-M^{kl}-M^{lk}+M^{ll})}\\
    &\hspace{5pt}\times e^{\ii(M^{km}-M^{lm})(\hat{a}_m^\dag \hat{a}_m +\frac{1}{2})}\hat{a}_k^\dag\hat{a}_l\,,
\end{split}
\end{align}
and the corresponding conjugate relations, which follow from~\eqref{eq:action-U_W-on-E}. Combining transformations of these types, the expectation value on the states~\eqref{eq:generalized-gaussian-states-ex} of any polynomial of creation and annihilation operators can be brought to the standard form of linear combinations of
\begin{equation}
    \braket{0|\Ug{S} \hat{\mathbf{x}}_{i_1}\cdots\hat{\mathbf{x}}_{i_n}|0}\,.
    \label{eq:standard-form-gaussian-ex}
\end{equation}
\end{example}

\section{Efficient evaluation of expectation values in standard form}
\label{sec:efficient-computation-global}
Generalized group-theoretic coherent states will only be useful as variational families if we can efficiently evaluate expectation values $\braket{\psi|\hat{\mathcal{O}}|\psi}$. In the previous section, we have shown that any such expectation value can be reduced to the standard form~\eqref{eq:expectation-U-Xi}. To evaluate this standard form, we need to be able to compute its building blocks of the form
\begin{equation}
    \braket{\mu|\,\Ug{\M}\,\hat{\X}_{i_1}\cdots\hat{\X}_{i_n}|\mu}\,.
    \label{eq:expectation-U-Xi-eff}
\end{equation}
In this section, we will discuss how to compute~\eqref{eq:expectation-U-Xi-eff} efficiently and thereby evaluate arbitrary expectation values from the standard form \eqref{eq:expectation-U-Xi}.

\subsection{BCH decomposition}
\label{sec:efficient-computation}
Computing~\eqref{eq:expectation-U-Xi-eff} can be achieved by performing a normal ordered Baker-Campbell-Hausdorff decomposition, also known as Gauss decomposition, of the group unitary $\Ug{\M}$ that appears in it. Let us assume that $\Ug{\M}$  can be written as an exponential of algebra elements. We therefore have
\begin{align}
    \Ug{\M}&=\exp\left(\sum_{\al\in\Delta_+}K^{\al}_+\hat{E}_{\al} + K_0^\I \hat{H}_\I + \sum_{\al\in\Delta_+}K^\al_-\hat{E}_{-\al}\right)\,,
    \label{eq:U-as-exp}
\end{align}
where we have used that the algebra operators $\hat{\X}_i$ can be decomposed on the basis $\hat{H}_\I,\hat{E}_\al$ and we have introduced the corresponding complex coefficients $K_0^\I$, $K^\al_\pm$. We would like to split the exponential appearing in~\eqref{eq:U-as-exp} into the product of three terms and rewrite $\Ug{\M}$ as 
\begin{align}
    \Ug{\M}=\hat{T}_+\, \hat{T}_0\, \hat{T}_-\,,
    \label{eq:BCH}
\end{align}
where $\hat{T}_\pm$ and $\hat{T}_0$ are operators of the forms
\begin{align}
\hat{T}_\pm=\exp\left(\sum_{\al\in\Delta_+}\!A^{\al}_\pm\,\hat{E}_{\pm\al}\right),\,\,
\hat{T}_0=\exp\left(A_0^\I\,\hat{H}_\I\right),
\end{align}
for some appropriate choice of the coefficients $A_0^\I$, $A^\al_\pm$.

The specific functional dependence of $A_0^\I$ and $A^\al_\pm$ on $K_0^\I$ and $K^\al_\pm$ and the extent to which it can be calculated analytically will depend on the given choice of the group $\mathcal{G}$. However, let us point out that the decomposition~\eqref{eq:BCH} only depends on the abstract group and algebra properties and not on the specific choice of representation. It may therefore be convenient to perform such decomposition working in a smaller representation than the one of the physical system, \eg the fundamental or adjoint representation.

Once the decomposition~\eqref{eq:BCH} of $\Ug{\M}$ has been performed the computation of the expectation value~\eqref{eq:expectation-U-Xi-eff} becomes relatively straightforward. Indeed, one can commute $\hat{T}_-$ to the right of the algebra operators $\hat{\X}_{i_1}\cdots\hat{\X}_{i_n}$ just giving rise to new linear combinations of algebra operators. To do this one needs a relation analogous\footnote{Formula~\eqref{eq:commute-left-non-unitary} and the form of matrix $\mathbf{R}$ can be derived in the same way as~\eqref{eq:adjoint-representation} and~\eqref{eq:commute-left} as explained in footnote~\ref{foot:adjoint-rep} and Appendix~\ref{app:adjoint-rep}. Note that we have here the quantity $\hat{T}_-$ (instead of $\Ug{\M}$) which is not a unitary operator, but is still the exponential of complex combinations of algebra elements.} to equation~\eqref{eq:commute-left}, \ie
\begin{equation}
    \hat{T}_- \,\hat{\X}_i \, =\mathbf{R}_i^j \,  \hat{\X}_j \,\hat{T}_-\,.
    \label{eq:commute-left-non-unitary}
\end{equation}
In this way, one reduces~\eqref{eq:expectation-U-Xi-eff} to the form
\begin{align}
\begin{split}
    &\mathbf{R}_{i_1}^{j_1}\cdots\mathbf{R}_{i_n}^{j_n}\,\braket{\mu|\hat{T}_+\hat{T}_0 \,\hat{\X}_{j_1}\cdots\hat{\X}_{j_n} \,\hat{T}_- |\mu}\\
    &\hspace{40pt}=e^{ \ii A_0^\I \mu_\I} \, \mathbf{R}_{i_1}^{j_1}\cdots\mathbf{R}_{i_n}^{j_n}\,\braket{\mu|\hat{\X}_{i_1}\cdots\hat{\X}_{i_n} |\mu} \,,
\end{split}
    \label{eq:expectation-final}
\end{align}
where we used that the lowest weight vector $\ket{\mu}$ is left-invariant by $\hat{T}_-$ on the right, right-invariant by $\hat{T}_{+}$ on the left and is an eigenstate with eigenvalue $\ii\mu_\I$ of the operators $\hat{H}_\I$ that appear in $\hat{T}_0$. Let us stress again that the eigenvalues $\mu_\I$ are the only object in this derivation that depends on the choice of representation that we are using.

The information on the group element $\M$ appearing in the original expression~\eqref{eq:expectation-U-Xi-eff} is contained in the linear coefficients $\mathbf{R}_i^j$ (which will depend on $A_-^\al$) and in the coefficients $A_0^\I$ that appear in the first factor of~\eqref{eq:expectation-final}. The factor $\braket{\mu|\hat{\X}_{i_1}\cdots\hat{\X}_{i_n} |\mu}$ is instead independent of $\M$ and thus needs to be computed only once. This can be done using the standard algebra commutation relations.

\begin{example}[BCH for spin-$\frac{1}{2}$ coherent states]
As $\mathcal{U}(g)$ is always a tensor product over individual spin degrees of freedom, we can evaluate the standard form of the expectation value for each one individually. We thus consider
\begin{align}
    \braket{\downarrow\!\!|e^{\ii K^i\sigma_i}\hat{\sigma}_{i_1}\dots \hat{\sigma}_{i_n}|\!\!\downarrow}\,.
\end{align}
The BCH decomposition of $e^{\ii K^i\sigma_i}$ is well-known~\cite{arecchi1972atomic,ringel2013dynamical} and explicitly given by
\begin{equation}
    e^{K_+ \hat{\sigma}_++\ii\frac{K_0}{2}\hat{\sigma}_3 - K_+^* \hat{\sigma}_-}=e^{A_+ \hat{\sigma}_+}e^{\frac{A_0}{2}\hat{\sigma}_3}e^{A_- \hat{\sigma}_-}\,,
    \label{eq:su2-decomposition-example}
\end{equation}
where the respective coefficients are given by
\begin{align}
    A_0&=-2\log \left(\cos \varphi - \tfrac{1}{2} K_0 \tfrac{\sin \varphi}{\varphi}\right) \\
    A_+&=A_-^*=-\ii K_+ \tfrac{\sin \varphi}{\varphi} {\left(\cos \varphi - \tfrac{1}{2} K_0 \tfrac{\sin \varphi}{\varphi}\right)}^{-1}\,,
\end{align}
with $\varphi=\sqrt{{|K_+|}^2+\frac{1}{4}K_0^2}$. To find the equivalent of~ \eqref{eq:expectation-final}, we can use~\eqref{eq:su2-adjoint-example} to deduce $e^{A_-\hat{\sigma}_-} \;\hat{\sigma}_i=\mathbf{R}_{ij} \,\hat{\sigma}_j \; e^{A_-\hat{\sigma}_-}$ with
\begin{align}
  \footnotesize \mathbf{R}=\left(\begin{array}{ccc} 1-\frac{1}{4}A_-^2 & \frac{\ii}{4}A_-^2 & \frac{1}{\sqrt{2}}A_- \\
  \frac{\ii}{4}A_-^2 & 1+\frac{1}{4}A_-^2 & -\frac{\ii}{\sqrt{2}}A_- \\ -\frac{1}{\sqrt{2}}A_- & \frac{\ii}{\sqrt{2}}A_- & 1 \end{array} \right) \,.
\end{align}
Combining these results, we thus find
\begin{align}
    \braket{\downarrow\!\!|e^{\ii K^i \hat{\sigma}_i}\, \!\hat{\sigma}_{i_1}\!\!\cdots\hat{\sigma}_{i_n}|\!\!\downarrow}=e^{s A_0}\; \mathbf{R}_{i_1 j_1}\!\!\cdots \mathbf{R}_{i_n j_n} \braket{\downarrow\!\!|\hat{\sigma}_{j_1}\!\!\cdots\hat{\sigma}_{j_n}|\!\!\downarrow}
\end{align}
with $s=-\frac{1}{2}$ for spin-$\frac{1}{2}$, which generalizes easily to larger spin.
\end{example}

\begin{example}[BCH for bosonic Gaussian states]
To evaluate~\eqref{eq:standard-form-gaussian-ex} via BCH we first can decompose the unitary as $\Ug{S}=\Ugd{u}\Ug{T}$, where $\bra{0}\Ugd{u}=e^{-\ii\theta}\bra{0}$ and
\begin{equation}
    \Ug{T}=\exp \left( (K_+)_{kl} \,\ii\,\hat{a}^\dag_k \,\hat{a}^\dag_l + (K_+^*)_{kl} \, \:\ii\,\hat{a}_k \hat{a}_l\right)\,,
\end{equation}
for a suitable $K_+$. For this type of unitary the decomposition $\Ug{T}=\hat{T}_+ \hat{T}_0 \hat{T}_-$ is known analytically~\cite{windt_local_2020}. Using this decomposition one can obtain the final result
\begin{equation}
    \braket{0|\,\Ug{S} \hat{\mathbf{x}}_{i_1}\cdots\hat{\mathbf{x}}_{i_n}|0}= r_0 \mathbf{R}_{i_1 j_1 }\cdots \mathbf{R}_{i_n j_n}\! \braket{0| \hat{\mathbf{x}}_{j_1}\cdots\hat{\mathbf{x}}_{j_n}|0},
\end{equation}
where $r_0$ is given by
\begin{align}
    r_0=e^{-\frac{\ii}{4}\mathrm{tr}(\Omega \log\sqrt{S^{^\intercal}\! S}S^{-1})}\det(\id-4 A_+ A_+^*)^{\frac{1}{4}}\,,
\end{align}
and $\mathbf{R}$ is the $2N\times 2N$ matrix
\begin{equation}
    \mathbf{R}=\left(\begin{array}{cc} \id-A_+^* & -\ii A_+^* \\ -\ii A_+^* & \id+A_+^* \end{array}\right)\,.
\end{equation}
The matrix $A_+$ can be derived analytically from $S$ according to~\eqref{eq:gaussian-A-plus}. See Appendix~\ref{app:ex-gaussian} for a more detailed derivation.
\end{example}

\subsection{Time evolution of the BCH decomposition}
In the previous section, we showed how to compute~\eqref{eq:expectation-U-Xi-eff} which required a normal ordered Baker-Campbell-Hausdorff decomposition of $\Ug{\M}$ for every $\M$. For many standard Lie groups, the needed formulas already exist in the literature. However, this decomposition can also be computed by solving a corresponding set of differential equations. This approach can be used if the respective closed analytical formulas are not known or difficult to implement and is especially convenient in settings where one performs time evolution.

Time evolution is an important application of generalized group-theoretic coherent states, where one uses them to simulate the dynamics of quantum systems, either in real time or imaginary time. A similar setting is the one where one applies gradient descent methods to our family of states. In all these applications one has the need to compute a certain set of expectation values at each time step of the evolution, then update the state to a new one which is (theoretically) infinitesimally close and repeat the procedure. Therefore, one is required to calculate the decomposition~\eqref{eq:BCH} at a series of subsequent time steps as $\M$ evolves as a function of time (more precisely, $\M$ is a function of the variational parameters which in turn evolve as functions of time). In these settings, it would be useful if one could compute the BCH decomposition for $\Ug{\M(t+dt)}$ based on the decomposition of $\Ug{\M(t)}$ at the previous time step, instead of having to compute it from scratch at each step. We will now show how this can be done. As already mentioned above, this will also lead to a general method for computing~\eqref{eq:BCH}, that, although not always the most efficient, can be useful in cases where a closed formula is not available.

Let us assume that $\Ug{\M(t)}$ can be written as 
\begin{align}
    \Ug{\M(t)}=e^{K^i(t)\,\hat{\X}_i}\label{eq:U-time-dependent}
\end{align}
and that we want to decompose it as
\begin{align}
    \Ug{\M(t)}=\hat{T}_+(t)\,\hat{T}_0(t)\,\hat{T}_-(t)\,,
    \label{eq:bch-time-dependent}
\end{align}
where $\hat{T}_-(t)$, $\hat{T}_0(t)$ and $\hat{T}_+(t)$ are operators of the forms
\begin{subequations}
\begin{align}
\hat{T}_-(t)&=e^{\sum_{\al\in\Delta_+}\!A^{\al}_-(t)\,\hat{E}_{-\al}}\,,\\
\hat{T}_0(t)&=e^{A_0^\I(t) \,\hat{H}_\I}\,,\\
\hat{T}_+(t)&=e^{\sum_{\al\in\Delta_+}\!A^{\al}_+(t)\,\hat{E}_{\al}}\,.
\end{align}
\end{subequations}

We now take the time derivative of $\Ug{\M(t)}$ and multiply it by $\Ugm{\M(t)}$. From~\eqref{eq:U-time-dependent}, we have
\begin{align}
\begin{split}
    &\Ugm{\M(t)}\frac{d}{dt} \Ug{\M(t)}\\
    &\hspace{20pt}=\int_0^1 \! d\tau \: e^{-\tau K^j(t)\,\hat{\X}_j} \left[\frac{d}{dt} K^i(t)\,\hat{\X}_i\right] e^{\tau K^j(t)\,\hat{\X}_j}\label{eq:taos-integral}
    \end{split}\\
    &\hspace{20pt}=\left[\int_0^1 \! d\tau \: e^{\tau\,\mathrm{ad}(K(t))}\right]^i_j \frac{d}{dt} K^j(t)\,\hat{\X}_i\\
    &\hspace{20pt}=\left[\mathrm{ad}(K(t))^{-1}\left(e^{\mathrm{ad}(K(t))}-\id \right)\right]^i_j \: \frac{d}{dt} K^j(t)\,\hat{\X}_i
    \label{eq:time-evolution-left}
\end{align}
where $\mathrm{ad}(K(t))$ represents the matrix
\begin{equation}
    \left[\mathrm{ad}(K(t))\right]^i_j = K^k(t) \, c_{kj}^i \,,
\end{equation}
similarly to what explained in footnote~\ref{foot:adjoint-rep}.
For the expression used in~\eqref{eq:taos-integral} see, \eg the appendix of~\cite{shi2018variational}.

From~\eqref{eq:bch-time-dependent}, we have
\begin{align}
\begin{split}
    &\Ugm{\M(t)}\frac{d}{dt} \Ug{\M(t)}\\
    &\hspace{5pt}= \hat{T}_-(t)^{-1} \hat{T}_0(t)^{-1} \Big[\sum_{\al\in\Delta_+}\!\!d^{\al}_+(t)\hat{E}_{\al}\Big] \hat{T}_0(t)\, \hat{T}_-(t)\\
    &\hspace{5pt}+\hat{T}_-(t)^{-1} \left[d_0^\I(t) \hat{H}_\I\right] \hat{T}_-(t) + \Big[\sum_{\al\in\Delta_+}\!d^{\al}_-(t)\hat{E}_{-\al}\Big]\,.
    \label{eq:time-evolution-right}
\end{split}
\end{align}
The coefficients $d_0^\I(t)$ and $d^\al_\pm(t)$ are defined by\footnote{Note that in general $d^\al_\pm(t)\neq\frac{d}{dt}A^\al_\pm(t)$, because not all $\hat{E}_\al$ commute among themselves.} 
\begin{subequations}
\begin{align}
\frac{d}{dt} \hat{T}_-(t)&=\hat{T}_-(t)\Big[\sum_{\al\in\Delta_+}\!\!d^{\al}_-(t)\hat{E}_{-\al}\Big]\,,\label{eq:derivative-T-first}\\
\frac{d}{dt} \hat{T}_0(t)&=\hat{T}_0(t)\left[d_0^\I(t) \hat{H}_\I\right]\,,\\
\frac{d}{dt} \hat{T}_+(t)&=\hat{T}_+(t)\Big[\sum_{\al\in\Delta_+}\!d^{\al}_+(t)\hat{E}_{\al}\Big]\,.\label{eq:derivative-T-last}
\end{align}
\end{subequations}
By applying relations analogous to~\eqref{eq:commute-left-non-unitary}, equation~\eqref{eq:time-evolution-right} can be brought to the form of a linear combination of the algebra basis operators $\hat{\X}_i$, similarly to~\eqref{eq:time-evolution-left}. 

Finally, comparing these algebra elements, one can write $d_0^\I(t)$ and $d^\al_\pm(t)$ as functions of $\frac{d}{dt}K_0^\I$ and $\frac{d}{dt}K^\al_\pm$ and of $A_0^\I(t)$ and $A^\al_\pm(t)$. More precisely, equating~\eqref{eq:time-evolution-right} and~\eqref{eq:time-evolution-left} leads to
\begin{align}
\begin{split}
    &\mathbf{M}[A_0(t),A_-(t)] \,\left(\begin{array}{c} d^\al_-(t) \\ d^\I_0(t)  \\ d^\al_+(t)\end{array}\right)\\
    &\hspace{0pt}=\left[\mathrm{ad}(K(t))^{-1}\left(e^{\mathrm{ad}(K(t))}-\id \right)\right]\frac{d}{dt}\left(\begin{array}{c} K^\al_-(t) \\ K^\I_0(t)  \\ K^\al_+(t)\end{array}\right)\,,
\end{split}\label{eq:diff-eq}
\end{align}
where $\mathbf{M}[A_0(t),A_-(t)]$ is a matrix of the dimension of the algebra, that depends on $A_0(t)$ and $A_-(t)$ through the adjoint representation of the corresponding group elements, and which we need to invert.

Note that here the derivatives $\frac{d}{dt}K(t)$ depend only on how we update the variational parameters at the given time step and how this update influences $\M(t)$. We therefore assume them to be known. Similarly, the quantities $K(t)$, $A_0(t)$ and $A_-(t)$ depend only on the group element $\M(t)$ and on its BCH decomposition at the current time step. Having found $d_0^\I(t)$, $d^\al_\pm(t)$ from equation~\eqref{eq:diff-eq}, we can then integrate equations~\eqref{eq:derivative-T-first} to~\eqref{eq:derivative-T-last} for one time step to obtain the BCH decomposition~\eqref{eq:bch-time-dependent} at time $t+dt$.

If instead we just want to compute the Baker-Campbell-Hausdorff decomposition for a fixed group transformation of the form~\eqref{eq:U-as-exp}, we can write $K^i(t)=t K^i$ and integrate from $t=0$ to $t=1$ the corresponding differential equations~\eqref{eq:derivative-T-first} to~\eqref{eq:derivative-T-last} as described in this section to obtain the desired decomposition~\eqref{eq:BCH}.

\section{Summary and Outlook}\label{sec:summary}
In this manuscript, we have introduced \emph{generalized group-theoretic coherent states} as a new family of pure quantum states. This family is defined on top of the well-known \emph{Gilmore-Perelomov group-theoretic coherent states} by applying an additional unitary $\Ung{\W}$. There exist many examples of group-theoretic coherent states, defined by different choices of Lie groups and representations, and this makes our construction quite general and applicable in various contexts. 

The transformation $\Ung{\W}$ is defined as the exponential of a quadratic expression in the so-called Cartan subalgebra operators $\hat{H}_a$. This introduces quantum correlations not contained in traditional group-theoretic coherent states, thus allowing the treatment of problems beyond mean-field. The dynamics of regular group-theoretic coherent states correspond to the group-theoretic version of semi-classical Landau-Lifshitz (LL) equations for $\mathrm{SU}(2)$ spin models \cite{lakshmanan2011fascinating}. Our new class of wavefunctions allows in this sense to go beyond semi-classical dynamics. In particular, we expect generalized coherent states to be suitable for systems with interacting Hamiltonians containing terms also quadratic in Cartan operators. For these, it will be interesting to explore whether the many exact theoretical results that have been proven for the Landau-Lifshitz equations, such as existence of solitons in 1d, will be be robust to going beyond the LL factorizable wavefunction ansatz.
We further emphasized that generalized group theoretic states are particularly powerful when we want to correlate different types of degrees of freedom (\eg spins, bosons, fermions) in composite systems, as the transformation $\Ung{\W}$ can be used to entangle them by including Cartan generators of different types.

While going beyond coherent states, we showed in section~\ref{sec:efficient-computation-global} that generalized coherent states still allow for an efficient evaluation of generic expectation values. We stress, however, that computing the overlap $\braket{\psi|\tilde\psi}$ between two arbitrary generalized group-theoretic coherent states $\ket{\psi}$ and $\ket{\tilde\psi}$ remains in general a hard task.

We gave two key examples of how our construction can be applied in different settings, namely for spin-$\frac{1}{2}$ coherent states and bosonic Gaussian states. However, the range of applications of our proposal is by no means limited to these examples: they can be extended, combined or complemented in many ways. The $\mathrm{SU}(2)$ construction can, for instance, be extended to higher spin representations, for example to atomic coherent states~\cite{arecchi1972atomic} obtaining so-called \emph{spin squeezed states}~\cite{kitagawa_squeezed_1993}.
The Gaussian state construction can be repeated for \emph{fermionic} Gaussian states, as sketched in Appendix~\ref{app:ex-fermionic-gaussian}.

It is also straightforward to apply the described generalization to more elaborate Lie groups and algebras~\cite{mathur_sun_2002,galitski2011quantum}. This is particularly useful as many lattice systems can be described as an $\mathrm{SU}(N)$ problem, where $N$ is the dimension of the Hilbert space at a site \cite{batista2004algebraic,davidson2015s}. Our approach can thus be used to study dynamics with variational states that have non-trivial entanglement utilizing this $\mathrm{SU}(N)$ perspective. Finally, a further interesting possibility is that of defining $\Ung{\W}$ in terms of a choice of Cartan subalgebra different from the one with respect to which the reference state $\ket{\mu}$ is a lowest weight state, which can be done for non-compact Lie groups, such as $\mathrm{Sp}(2N,\mathbb{R})$ for bosonic Gaussian states.

We currently restricted ourselves to semi-simple Lie groups, as those are the ones studied systematically in mathematical physics and for which the construction of Cartan subalgebra and root system is fully understood. While this enabled us to present a systematic framework of generalized group-theoretic coherent states, we know that in special cases we can follow the same philosophy also for Lie groups that are not semi-simple. The most prominent example is the Heisenberg group associated to displacement operators for bosonic degrees of freedom, which plays the key role in the definition of regular bosonic coherent states. It will be an interesting exercise to explore the full extent to which this group can be incorporated in our formalism and consider whether the same can be done for other non-semi-simple groups.

Some of the examples discussed above have already been proposed and studied~\cite{shi2018variational}. A few of them already have a history of successful applications. For example, by choosing a fermionic number operator $\hat{n}_{\mathrm{f}}$ and a bosonic quadrature operator $\hat{p}=\frac{\ii}{\sqrt{2}}(\hat{a}^\dag-\hat{a})$ as Cartan-type generators we obtain a $\Ung{M}$ that corresponds to the well-known Lang-Firsov Polaron transformation~\cite{lang_ij_kinetic_1963}, often used for correlated boson-fermion systems. However, the presented framework can lead to a whole spectrum of new generalizations which we believe can be of great interest.

In terms of concrete applications, we believe that interesting developments can come from two directions. First, as our states are particularly amenable to being produced in common experimental implementations \emph{and} their expectation values can be computed efficiently by classical computation, they provide an ideal setting for benchmarking experimental set-ups and quantum computer prototypes. Second, they can be applied as variational states to describe and understand ground state and dynamical properties of many quantum many-body systems. Some families of states that can be understood as generalized coherent states have already been successfully employed to perform both exact and variational calculations~\cite{Foss-Feig_2013_Noneq,ashida2019quantum,wang2019zero,shchadilova2016polaronic}, testifying to the large spectrum of potential applications of the construction.
In particular, they include systems that contain bosons or fermions or both, for which our construction allows to go beyond a Gaussian approach and also caters for the necessity of entangling the bosonic and fermionic sectors. One can also consider systems where a spin impurity is coupled to a bosonic, fermionic or spin bath, such as the paradigmatic Kondo~\cite{kondo_resistance_1964,florens_kondo_2006} and Bose polaron models~\cite{landau_effective_1948,frohlich_interaction_1952}. We can finally take in consideration pure spin problems for which tensor network methods do not give satisfactory results, \eg in higher dimensions.

Some specific systems of the types above for which we believe generalized coherent states would represent an interesting novelty include the case of fermions with bi-phonon coupling~\cite{kennes_transient_2017}, where the interaction is given by $\hat{H}_{\mathrm{e-ph}}=\sum_i \hat{Q}_i^{(\mathrm{f})} \hat{Q}_i^{(\mathrm{b})}$, where $\hat{Q}_i^{(\mathrm{f})}$ and $\hat{Q}_i^{(\mathrm{b})}$ are respectively fermionic and bosonic quadratic operators.
Of interest is also the case of the Jahn-Teller polaron~\cite{gunnarsson_alkali-doped_2004} where, after a Lee-Low-Pines transformation~\cite{lee_motion_1953}, the Hamiltonian takes the form $\hat{H}_{\mathrm{e-ph}}=\sum_i \hat{F}_i \hat{q}_i$. Here, the $\hat{q}_i$ are quadratures of a bosonic bath and the $\hat{F}_i$ are a set of fermionic operators realizing an $\mathfrak{su}(2)$ algebra, which could be described by generalized spin-$\frac{1}{2}$ coherent states.

\begin{acknowledgments}
TG and IC are supported by the Deutsche Forschungsgemeinschaft (DFG, German Research Foundation) under Germany’s Excellence Strategy – EXC-2111 – 39081486. LH acknowledges support by VILLUM FONDEN via the QMATH center of excellence (grant no.10059). TS acknowledges funding through NSFC 11974363. ED acknowledges funding through Harvard-MIT CUA, ARO grant number W911NF-20-1-0163, the National Science Foundation through grants No. OAC-1934714 and NSF EAGER-QAC-QSA:
Quantum Algorithms for Correlated Electron-Phonon System, award number 2222-206-2014111. IC acknowledges funding through ERC Grant QUENOCOBA, ERC-2016-ADG (grant no.742102).
\end{acknowledgments}

\appendix
\section{\texorpdfstring{Spin-$\frac{1}{2}$}{Spin-1/2} coherent states}\label{app:ex-su2}
In this appendix, we illustrate in more detail the construction of generalized group-theoretic coherent states in the case of spin-$\frac{1}{2}$ coherent states. In the following subsections, we follow the structure of the main body of the paper illustrating the construction step-by-step.

\subsection{Group-theoretic coherent states}\label{app:group-theoretic-coherent-states-ex-su2}
This example arises if we make the Lie group choice $\mathcal{G}=\mathrm{SU}(2)$ with Lie algebra $\mathfrak{g}=\mathfrak{su}(2)$.

We consider the fundamental representation, \ie the spin-$\frac{1}{2}$ representation. We represent group elements $g\in\mathcal{G}$ as unitary $2\times 2$ matrices $\mathcal{U}$ and algebra elements as $2\times 2$ traceless anti-Hermitian matrices $\hat{K}$. These matrices act on a $2$-dimensional Hilbert space $\mathcal{H}_\frac{1}{2}=\mathbb{C}^2=\mathrm{span}\{\ket{\uparrow},\ket{\downarrow}\}$.

We can express any algebra element $\hat{K}$ in the basis of Pauli matrices, \ie $\hat{K}=\ii K^i \,\hat{\sigma}_i$, with
\begin{equation}
    \hat{\sigma}_1=\left(\begin{array}{cc} 0 & 1 \\ 1 & 0 \end{array}\right), \; \hat{\sigma}_2=\left(\begin{array}{cc} 0 & -\ii \\
    \ii & 0 \end{array}\right), \hat{\sigma}_3=\left(\begin{array}{cc} 1 & 0 \\ 0 & -1 \end{array}\right)\,,
\end{equation}
and some real coefficients $K^i$. Any group element $\mathcal{U}$ can be written as the exponential $\mathcal{U}=e^{\ii K^i \hat{\sigma}_i}$. Consequently, we choose the basis
\begin{equation}
    \hat{\X}_1=\ii\hat{\sigma}_1, \quad \hat{\X}_2=\ii\hat{\sigma}_2, \quad \hat{\X}_3=\ii\hat{\sigma}_3\,,
\end{equation}
whose commutation relations~\eqref{eq:structure-constants} are well-known as $[\ii\hat{\sigma}_i,\ii\hat{\sigma}_j]=-2\epsilon_{ijk}\ii\hat{\sigma}_k$.

These relations can also be used to construct the adjoint representation, where equation~\eqref{eq:adjoint-representation} takes the form
\begin{equation}
    e^{-\ii K^i\hat{\sigma}_i}\, \left(\begin{array}{c} \hat{\X}_1 \\\hat{\X}_2 \\\hat{\X}_3 \end{array}\right) \,e^{\ii K^i \hat{\sigma}_i} = e^{-2 K^i \mathbf{L}_i}\, \left(\begin{array}{c} \hat{\X}_1 \\\hat{\X}_2 \\\hat{\X}_3 \end{array}\right)
    \label{eq:su2-adjoint-representation}
\end{equation}
with $\mathbf{L}_i$ are the $3\times 3$ antisymmetric matrices
\begin{align}
\footnotesize   \mathbf{L}_1\!=\!\left(\begin{array}{ccc} 0 & 0 & 0 \\ 0 & 0 & -1 \\ 0 & 1 & 0  \end{array}\right),\, \mathbf{L}_2\!=\!\left(\begin{array}{ccc} 0 & 0 & 1 \\ 0 & 0 & 0 \\ -1 & 0 & 0  \end{array}\right),\,\mathbf{L}_3\!=\!\left(\begin{array}{ccc} 0 & -1 & 0 \\ 1 & 0 & 0 \\ 0 & 0 & 0  \end{array}\right)\,.
\end{align}

Let us now examine more in detail the structure of the algebra $\mathfrak{su}(2)$. The maximal set of mutually commuting algebra operators is one dimensional, \ie the algebra has rank $\ell=1$. We can therefore choose a single operator $\hat{H}$ as basis of the Cartan subalgebra, which we choose to be $\hat{H}=\frac{1}{2}\hat{\X}_3= \frac{\ii}{2} \hat{\sigma}_3$ without loss of generality.

Corresponding to this choice, we can identify a single root pair composed of the positive root $\al=1$ and the associated negative root $-\al=-1$. The respective root space operators are
\begin{align}
    \hat{E}_{\pm\al}&=\hat{\sigma}_\pm=\frac{1}{2\sqrt{2}}(\hat{\sigma}_1\pm\ii\hat{\sigma}_2)=\frac{1}{2\sqrt{2}}(-\ii\hat{\X}_1\pm\hat{\X}_2)\,.
    \label{eq:root-space-operators-su2}
\end{align}
The relation~\eqref{eq:adjoint-action-H-on-E} then takes the form
\begin{equation}
    [\frac{\ii}{2} \hat{\sigma}_3,\hat{\sigma}_{\pm}]=\pm \ii \hat{\sigma}_{\pm}\,. 
\end{equation}
From~\eqref{eq:root-space-operators-su2}, we see that $\hat{E}_{\pm\al}$ are complex linear combinations of $\hat{\X}_i$ and are therefore not themselves operators of $\mathfrak{su}(2)$, as they are not anti-Hermitian. However, all algebra operators can be expressed as complex linear combinations of $\hat{H}=\frac{\ii}{2} \hat{\sigma}_3,\hat{E}_{+\al}=\hat{\sigma}_+,\hat{E}_{-\al}=\hat{\sigma}_-$.

The weight vectors of this representation are the basis vectors $\ket{\downarrow}$ and $\ket{\uparrow}$, as they are eigenvectors of $\hat{H}=\frac{\ii}{2} \hat{\sigma}_3$. In particular, the lowest weight vector is $\ket{\downarrow}$, as it is annihilated by the negative root operator, \ie $\hat{E}_{-\al}\ket{\downarrow}=\hat{\sigma}_{-}\ket{\downarrow}=0$. As discussed in section~\ref{sec:group-theoretic-coherent-states-global}, this state will be chosen for the role of reference state in the definition of group-theoretic coherent states, \ie $\ket{\phi}=\ket{\downarrow}$. This leads to the definition of spin-$\frac{1}{2}$ coherent states as $\mathcal{U}\ket{\downarrow}$, \ie
\begin{equation}
    \mathcal{M}_{\mathrm{SU}(2)}=\{ e^{\ii K^i \hat{\sigma}_i}\ket{\downarrow} \st K\in\mathbb{R}^3\}\,.
\end{equation}

Let us first note that the set $\mathcal{M}_{\mathrm{SU}(2)}$ includes all states of $\mathcal{H}_\frac{1}{2}=\mathbb{C}^2$ with unit norm. Second, for any vector $\ket{\phi}\in\mathcal{H}_\frac{1}{2}=\mathbb{C}^2$, there exists a choice of Cartan subalgebra and root ordering such that $\ket{\phi}$ is the lowest weight state.

A less trivial structure is obtained if instead of considering a single spin system, we consider a set of $N$ spins, described by the Hilbert space
\begin{equation}
    \mathcal{H}={\left(\mathcal{H}_\frac{1}{2}\right)}^{\otimes N}\,.
\end{equation}
Then we can choose as group the product of $N$ spin-$\frac{1}{2}$ representations of $\mathrm{SU}(2)$, each acting on one of the spins. The corresponding algebra will then be the sum of $N$ copies of $\mathfrak{su}(2)$. It can be expressed in terms of anti-Hermitian linear combinations of the operators
\begin{align}
    \hat{H}_k=\frac{\ii}{2}\hat{\sigma}_3^k\,, \quad \hat{E}_{\pm\al}^k=\hat{\sigma}_\pm^k\,,
    \label{eq:su2-algebra}
\end{align}
where the index $k=1,\dots,N$ refers to the spin on which the operators act. The Cartan subalgebra will be composed of the $N$ operators $\hat{H}_k=\frac{\ii}{2} \hat{\sigma}^k_3$, one for each spin $k$. In what follows we will consider this system of $N$ spin-$\frac{1}{2}$ degrees of freedom.

\subsection{Generalized family of states}\label{app:generalized-family-ex-su2}
We will now apply the construction introduced in section~\ref{sec:generalized-family}. Considering the Cartan subalgebra defined by the operators $\hat{H}_k$ in equation~\eqref{eq:su2-algebra}, the unitary operator~\eqref{eq:definition-U_W} takes the form 
\begin{equation}
    \Ung{M}=\exp \left(-\frac{i}{8}M_{kl}\,\hat{\sigma}_3^k \hat{\sigma}^l_3\right)\,,
\end{equation}
for any given $N\times N$ real symmetric matrix $M$. We see that the operator $\Ung{M}$ encodes correlations between the different spins.

Consequently, the \emph{generalized} $\mathrm{SU}(2)$ spin-$\frac{1}{2}$ coherent states take the form 
\begin{align}
    \ket{\psi(K_1,K_2,M)}&=\Ug{K_1}\,\Ung{M}\,\Ug{K_2}\ket{\downarrow}\,,\label{eq:generalized-coherent-states-su2}
\end{align}
where we recall that the group unitaries are defined as
\begin{equation}
    \Ug{K}=\exp\left(\ii K^{i,k} \hat{\sigma}^k_i\right)\,,
\end{equation}
with the coefficients $K^{i,k}$ taking values for $i=1,2,3$ and for each spin $k=1,\dots,N$.

All observables, \ie all Hermitian operators, can be written as polynomials of Pauli matrices and are therefore polynomials of algebra operators. To compute the expectation value of these observables on states~\eqref{eq:generalized-coherent-states-su2}, one needs to use the formulas
\begin{align}
    \hat{\sigma}_3^k \,\Ung{M}&=\Ung{M}\,\hat{\sigma}_3^k\,,\\
    \hat{\sigma}_\pm^k \,\Ung{M}&=\Ung{M}\,e^{-\frac{\ii}{2}M_{kk}}e^{\pm\frac{\ii}{2}M_{kl}\hat{\sigma}_3^l}\hat{\sigma}_\pm^k\,,
\end{align}
corresponding to~\eqref{eq:action-U_W-on-H} and~\eqref{eq:action-U_W-on-E}, and
\begin{equation}
    \left(\begin{array}{c} \hat{\sigma}_1^k \\\hat{\sigma}_2^k \\\hat{\sigma}_3^k \end{array}\right) \,\Ug{K} = \Ug{K}\, e^{-2 K^{i,k} \mathbf{L}_i}\, \left(\begin{array}{c} \hat{\sigma}_1^k \\\hat{\sigma}_2^k \\\hat{\sigma}_3^k \end{array}\right)\,,
\end{equation}
which corresponds to~\eqref{eq:commute-left} and can be derived immediately from~\eqref{eq:su2-adjoint-representation}.

Using these relations repeatedly one can commute all the operators $\Ug{K}$ and $\Ung{M}$to the left, which appear in the expectation value
\begin{align}
    &\braket{\psi(K_1,K_2,M)|\hat{\sigma}_{i_1}^{k_1}\cdots\hat{\sigma}_{i_n}^{k_n}|\psi(K_1,K_2,M)}
\end{align}
and combine them together with $\Ugd{K_1}$ and $\Ungd{M}$ coming from the bra vector to yield identities. What is left will be of the form of linear combinations of
\begin{equation}
    \braket{\downarrow\!|\Ug{K}\, \hat{\sigma}_{i_1}^{k_1}\cdots\hat{\sigma}_{i_n}^{k_n}|\!\downarrow}\,,
    \label{eq:su2-expectation-value}
\end{equation}
which we will show how to evaluate next and where $\Ug{K}$ is the combination of all the remaining group unitaries.

\subsection{Efficient computation of expectation values in standard form}
\label{app:efficient-computation-ex-su2}
We would like to compute quantities of the form of~\eqref{eq:su2-expectation-value}. Let us note that the group transformation $\Ug{K}$ appearing in such expression factorizes into unitaries acting locally on each site. As the operators $\hat{\sigma}_i^k$ are also all local, the problem reduces to a product of single site expectations of the type
\begin{equation}
    \braket{\downarrow\!|e^{\ii K^i \hat{\sigma}_i}\, \hat{\sigma}_{i_1}\cdots\hat{\sigma}_{i_n}|\!\downarrow}\,.
    \label{eq:su2-onsite-expectation}
\end{equation}
It is clear that computing~\eqref{eq:su2-onsite-expectation} only involves simple linear algebra of $2\times 2$ matrices, and can therefore be done efficiently without necessarily exploiting the techniques described in section~\ref{sec:efficient-computation}. Nonetheless, we will show how this would be done to illustrate the technique. Furthermore, the derived result can be equally applied to the case of higher spin representations, where the matrix algebra would become more cumbersome. We write the group operator appearing in~\eqref{eq:su2-onsite-expectation} as $\exp(K_+ \hat{\sigma}_++\ii\frac{K_0}{2}\hat{\sigma}_3 - K_+^* \hat{\sigma}_-)$ and decompose it as
\begin{equation}
    e^{K_+ \hat{\sigma}_++\ii\frac{K_0}{2}\hat{\sigma}_3 - K_+^* \hat{\sigma}_-}=e^{A_+ \hat{\sigma}_+}e^{\frac{A_0}{2}\hat{\sigma}_3}e^{A_- \hat{\sigma}_-}\,.
    \label{eq:su2-decomposition}
\end{equation}
By computing explicitly the matrix exponentials in this $2\times 2$ representation and comparing the two sides of~\eqref{eq:su2-decomposition} one finds~\cite{arecchi1972atomic}
\begin{align}
    A_0&=-2\log \left(\cos \varphi - \frac{1}{2} K_0 \frac{\sin \varphi}{\varphi}\right) \\
    A_+&=A_-^*=-\ii K_+ \frac{\sin \varphi}{\varphi} {\left(\cos \varphi - \frac{1}{2} K_0 \frac{\sin \varphi}{\varphi}\right)}^{-1}\,,
\end{align}
with $\varphi=\sqrt{{|K_+|}^2+\frac{1}{4}K_0^2}$. Note that this decomposition remains valid for any representation of the group $\mathrm{SU}(2)$, \ie we can replace $\hat{\sigma}_i$ with the operators $\hat{S}_i$ of larger spins. 

In the last step, we need to bring~\eqref{eq:su2-onsite-expectation} into the form~\eqref{eq:expectation-final} by commuting $e^{A_-\hat{\sigma}_-}$ to the right through all the $\hat{\sigma}_i$ operators. To do this, we observe that equation~\eqref{eq:commute-left-non-unitary} takes the form
\begin{equation}
    e^{A_-\hat{\sigma}_-} \;\hat{\sigma}_i=\mathbf{R}_{ij} \,\hat{\sigma}_j \; e^{A_-\hat{\sigma}_-}
    \label{eq:su2-commute-left-non-unitary}
\end{equation}
where
\begin{align}
\begin{split}
    \mathbf{R}&= e^{A_-(\ii\mathbf{L}_1+\mathbf{L}_2)}\\
    &=\left(\begin{array}{ccc} 1-\frac{1}{4}A_-^2 & \frac{\ii}{4}A_-^2 & \frac{1}{\sqrt{2}}A_- \\
    \frac{\ii}{4}A_-^2 & 1+\frac{1}{4}A_-^2 & -\frac{\ii}{\sqrt{2}}A_- \\
    -\frac{1}{\sqrt{2}}A_- & \frac{\ii}{\sqrt{2}}A_- & 1 \end{array} \right)
\end{split}
\end{align}
and where we used  in the second step that ${(\ii\mathbf{L}_1+\mathbf{L}_2)}^3=0$. In conclusion, we have the result
\begin{align}
\begin{split}
    &\braket{\downarrow|e^{\ii K^i \hat{\sigma}_i}\, \hat{\sigma}_{i_1}\cdots\hat{\sigma}_{i_n}|\downarrow}\\
    &\hspace{20pt}=e^{s A_0}\; \mathbf{R}_{i_1 j_1}\!\cdots \mathbf{R}_{i_n j_n}  \braket{\downarrow|\hat{\sigma}_{j_1}\cdots\hat{\sigma}_{j_n}|\downarrow}\,,
\end{split}
\end{align}
where $s=-\frac{1}{2}$ is the eigenvalue of $\frac{1}{2} \hat{\sigma}_3$ on $\ket{\downarrow}$. This easily generalizes to higher spin representations by replacing $s$ with the respective spin and $\hat{\sigma}_i$ with the respective $\hat{S}_i$.

\section{Bosonic Gaussian states}\label{app:ex-gaussian}
We review in further detail the example of bosonic Gaussian states, which is more elaborate than generalized spin-$\frac{1}{2}$ coherent states, as it involves the more complicated and non-compact Lie group $\mathrm{Sp}(2N,\mathbb{R})$. We restrict ourselves for simplicity to squeezing only, \ie without any coherent displacement. As before, we follow the structure of the main body of the paper illustrating the construction step-by-step.

\subsection{Group-theoretic coherent states}\label{app:group-theoretic-coherent-states-ex-gaussian}
We consider a system of $N$ bosonic modes, characterized by the position and momentum operators $\hat{q}_1,\dots,\hat{q}_N,\hat{p}_1,\dots,\hat{p}_N$. They are Hermitian operators which can also be expressed as $\hat{q}_k=\frac{1}{\sqrt{2}}(\hat{a}^\dag_k+\hat{a}_k)$ and $\hat{p}_k=\frac{\ii}{\sqrt{2}}(\hat{a}^\dag_k-\hat{a}_k)$, where $\hat{a}^\dag_k$ and $\hat{a}_k$ are the canonical creation and annihilation operators of the $k$-th mode. They satisfy the commutation relations
\begin{equation}
    [\hat{q}_k,\hat{q}_l]=[\hat{p}_k,\hat{p}_l]=0, \quad [\hat{q}_k,\hat{p}_l]=\ii\delta_{kl}\,.
    \label{eq:gaussian-commutators}
\end{equation}

Gaussian unitaries are defined as operators of the form $\mathcal{U}=e^{\hat{Q}}$, where $\hat{Q}$ is any anti-Hermitian homogeneous order 2 polynomial in the operators $\hat{q}_k,\hat{p}_k$. More precisely, if we group all the position and momentum operators into a single $2N$-dimensional vector $\hat{\mathbf{x}}=(\hat{q}_1,\dots,\hat{q}_N,\hat{p}_1,\dots,\hat{p}_N)^\intercal$, $\hat{Q}$ can be put in the form
\begin{equation}
    \hat{Q}=\frac{\ii}{2} \: \hat{\mathbf{x}}^\intercal h \hat{\mathbf{x}}\,,
\end{equation}
where $h$ is any $2N\times 2N$ real symmetric matrix. In principle $h$ could be any Hermitian matrix. However, using the commutation relations~\eqref{eq:gaussian-commutators} one can show that the anti-symmetric part of $h$ only contributes an imaginary c-number to $\hat{Q}$, therefore only a global phase to $\mathcal{U}$, in which we are not interested. So we can assume $h$ to be symmetric and real.

Gaussian states (also known as \emph{squeezed states}) are defined as the states obtained by acting with any Gaussian unitary on the Fock vacuum $\ket{0}$. Thus, Gaussian states are all of the form $e^{\hat{Q}}\ket{0}$ for any allowed $\hat{Q}$. Here, the vacuum is defined as the state annihilated by all annihilation operators, \ie $\hat{a}_k\ket{0}=0$, $\forall k$.

Bosonic Gaussian states defined in this way fit into the group-theoretic coherent states formalism described in section~\ref{sec:group-theoretic-coherent-states-global}. This is because the Gaussian operators $\mathcal{U}$ that we have defined give a unitary representation of the Lie group of real symplectic matrices\footnote{To be completely precise they are a unitary representation of the double cover of the group $\mathrm{Sp}(2N,\mathbb{R})$, known as the \emph{metaplectic group} $\mathrm{Mp}(2N,\mathbb{R})$.\label{foot:metaplectic-caveat}}
\begin{align}
    \mathrm{Sp}(2N,\mathbb{R})=\{S\in\mathrm{GL}(2N,\mathbb{R})\st S^\intercal\Omega S=\Omega\}\,,
\end{align}
where the matrix $\Omega$ is defined as
\begin{equation}
    \Omega=\left(\begin{array}{cc} 0 & \id_N \\ -\id_N & 0 \end{array} \right)\,.
    \label{eq:Omega}
\end{equation}
Similarly, the set of anti-Hermitian operators $\hat{Q}$ give a representation of the symplectic Lie algebra
\begin{align}
    \mathfrak{sp}(2N,\mathbb{R})&=\{K\in\mathfrak{gl}(2N,\mathbb{R})\st \Omega K+K^\intercal\Omega =0\}\,.
\end{align}
Indeed, for each matrix $K\in\mathfrak{sp}(2N,\mathbb{R})$, one can construct a symmetric matrix $h=\Omega K$ and the corresponding Hilbert space operator
\begin{equation}
    \hat{Q}(K)=\frac{\ii}{2} \:\hat{\mathbf{x}}^\intercal h \hat{\mathbf{x}}=\frac{\ii}{2}\: \hat{\mathbf{x}}^\intercal \Omega K \hat{\mathbf{x}}\,.
    \label{eq:gaussian-Q}
\end{equation}
Similarly, for any matrix $S\in\mathrm{Sp}(2N,\mathbb{R})$ that can be written as $S=e^K$ for some $K\in\mathfrak{sp}(2N,\mathbb{R})$, one can define the corresponding unitary
\begin{equation}
    \mathcal{U}(S)=\mathcal{U}(e^{K})=e^{\hat{Q}(K)}\,.
    \label{eq:gaussian-unitary}
\end{equation}
The operators $\mathcal{U}(S)$ constitute a group representation, in the sense that one can show that\footnote{As discussed in footnote~\ref{foot:metaplectic-caveat} they rigorously constitute a representation only of the double cover of the group. In practice this means that relation~\eqref{eq:group-multiplication-gaussian} may be valid only up to a sign. For more detail on how to compute such sign see~\cite{de_gossons_symplectic,hackl_in_prep}.}
\begin{equation}
    \mathcal{U}(S)\,\mathcal{U}(\tilde{S})=\mathcal{U}(S\tilde{S})\,.
    \label{eq:group-multiplication-gaussian}
\end{equation}

As in section~\ref{sec:group-theoretic-coherent-states-global}, the algebra operators $\hat{Q}$ defined in~\eqref{eq:gaussian-Q} can be expanded on a basis $\hat{\X}_i$. In this case, $\hat{Q}$ can be expanded as
\begin{align}
\begin{split}
    \hat{Q}&=A^{kl}\frac{\ii}{2}(\hat{q}_k\hat{q}_l+\hat{p}_k\hat{p}_l)+B^{kl}\frac{\ii}{2}(\hat{q}_k\hat{q}_l-\hat{p}_k\hat{p}_l)\\
    &\hspace{10pt}+ C^{kl}\frac{\ii}{2}(\hat{q}_k\hat{p}_l+\hat{p}_k\hat{q}_l)+D^{kl}\frac{\ii}{2}(\hat{q}_k\hat{p}_l-\hat{p}_k\hat{q}_l)\,,
\end{split}
\end{align}
for real symmetric $A^{kl}$, $B^{kl}$, $C^{kl}$ and real antisymmetric $D^{kl}$. Thus, all $\hat{Q}$ are real linear combinations of the operators
\begin{subequations}
\begin{align}
    \frac{\ii}{2}(\hat{q}_k\hat{q}_k+\hat{p}_k\hat{p}_k)&=\ii(\hat{a}_k^\dag \hat{a}_k +\frac{1}{2})\label{eq:Z-basis-gaussian-first}\\
    \frac{\ii}{2}(\hat{q}_k\hat{q}_l+\hat{p}_k\hat{p}_l)&=\frac{\ii}{2}(\hat{a}_k \hat{a}_l^\dag+\hat{a}_k^\dag \hat{a}_l ),\quad k< l\\
    \frac{\ii}{2}(\hat{q}_k\hat{p}_l-\hat{p}_k\hat{q}_l)&=\frac{1}{2}(\hat{a}_k \hat{a}_l^\dag -\hat{a}_k^\dag \hat{a}_l),\quad k< l\\
    \frac{\ii}{2}(\hat{q}_k\hat{q}_l-\hat{p}_k\hat{p}_l)&= \frac{\ii}{2}(\hat{a}_k\hat{a}_l+\hat{a}_k^\dag\hat{a}_l^\dag),\quad k\leq l\\
    \frac{\ii}{2}(\hat{q}_k\hat{p}_l+\hat{p}_k\hat{q}_l)&= \frac{1}{2}(\hat{a}_k\hat{a}_l-\hat{a}_k^\dag\hat{a}_l^\dag),\quad k\leq l \label{eq:Z-basis-gaussian-last}
\end{align}
\end{subequations}
which play the role of the operators $\hat{\X}_i$.

These can in turn be decomposed into combinations of Cartan subalgebra operators $\hat{H}_\I$ and root space operators $\hat{E}_\al$. More specifically, we can choose Cartan operators
\begin{equation}
    \hat{H}_k=\ii(\hat{a}_k^\dag \hat{a}_k +\tfrac{1}{2})\,,
    \label{eq:cartan-operators-gaussian}
\end{equation}
which leads to the root space operators
\begin{subequations}
\begin{align}
    \hat{E}_{+\al^{(k,l)}}&=\ii\hat{a}^\dag_k\hat{a}^\dag_l,&\hat{E}_{-\al^{(k,l)}}&=\ii\hat{a}_k\hat{a}_l,& k\leq l\\
    \hat{E}_{+\tilde{\al}^{(k,l)}}&=\hat{a}^\dag_k\hat{a}_l,&\hat{E}_{-\tilde{\al}^{(k,l)}}&=\hat{a}_k\hat{a}^\dag_l,&k<l
\end{align}
\end{subequations}
corresponding to the root vectors $\al^{\!(k,l)}_{\,\I}=(\delta_{ak}+\delta_{al})$ and $\tilde{\al}^{(k,l)}_{\,\I}=(\delta_{\I k}-\delta_{\I l})$. We see by inspection that all algebra operators $\hat{\X}_i$ as defined in equations~\eqref{eq:Z-basis-gaussian-first} to~\eqref{eq:Z-basis-gaussian-last} are complex linear combinations of these objects. The Fock vacuum $\ket{0}$ is the corresponding lowest weight state. Indeed, it is an eigenstate with eigenvalue $\frac{\ii}{2}$ of all Cartan subalgebra operators $\hat{H}_k$ and it is annihilated by all negative root space operators $\hat{E}_{-\al^{(k,l)}}\ket{0}=\hat{E}_{-\tilde{\al}^{(k,l)}}\ket{0}=0$.

We conclude that bosonic Gaussian states fulfil all the criteria to be identified as the group-theoretic coherent states for the group $\mathcal{G}=\mathrm{Sp}(2N,\mathbb{R})$, given its unitary representation in terms of bosonic operators described above.

\subsection{Generalized family of states}\label{app:generalized-family-ex-gaussian}
We now construct generalized bosonic Gaussian states following our definition in~\ref{sec:generalized-family}.  Based on~\eqref{eq:cartan-operators-gaussian}, we choose our Cartan subalgebra operators as
\begin{equation}
    \hat{H}_k=\ii(\hat{a}_k^\dag \hat{a}_k +\frac{1}{2})\,.
\end{equation} 
This leads to the non-Gaussian unitaries of the form
\begin{equation}
    \Ung{M}=\exp \left(-\frac{i}{2} M^{kl} (\hat{a}_k^\dag \hat{a}_k +\frac{1}{2})(\hat{a}_l^\dag \hat{a}_l +\frac{1}{2}) \right)
\end{equation}
for any $N\times N$ real symmetric matrix $M$. The \emph{generalized bosonic Gaussian states} are then defined as
\begin{equation}
    \ket{\psi(S_1,S_2,M)}=\Ug{S_1}\,\Ung{M}\,\Ug{S_2}\ket{0}\,,
\end{equation}
where $\Ug{S}$ are the Gaussian unitaries defined in~\eqref{eq:gaussian-unitary}. We recognize that these states constitute one of the classes of non-Gaussian states previously introduced in~\cite{shi2018variational}, which is not surprising as this construction heavily inspired us to define generalized group-theoretic coherent states in the prescribed way.

In this setting, the observables of interest will be polynomials in the operators $\hat{q}_k$ and $\hat{p}_k$, or equivalently in $\hat{a}_k^\dag$ and $\hat{a}_k$. As before, in order to compute expectation values of such observables, we need to commute them with unitaries of the types $\Ug{S}$ and $\Ung{M}$. This can be achieved thanks to the relations 
\begin{equation}
    \Ugd{S}\,\hat{\mathbf{x}}\,\Ug{S}=S\hat{\mathbf{x}}\,,
\end{equation}
which can be derived from~\eqref{eq:gaussian-commutators}, and
\begin{align}
\begin{split}
    \hspace{-7pt}\Ungd{M}\hat{a}_k\hat{a}_l \Ung{M}&=e^{-\frac{\ii}{2}(M^{kk}+M^{kl}+M^{lk}+M^{ll})}\\
    &\times e^{-\ii(M^{km}+M^{lm})(\hat{a}_m^\dag \hat{a}_m +\frac{1}{2})}\hat{a}_k\hat{a}_l\,,
\end{split}\\
\begin{split}
    \hspace{-7pt}\Ungd{M}\hat{a}_k^\dag\hat{a}_l \Ung{M}&=e^{-\frac{\ii}{2}(M^{kk}-M^{kl}-M^{lk}+M^{ll})}\\
    &\hspace{5pt}\times e^{\ii(M^{km}-M^{lm})(\hat{a}_m^\dag \hat{a}_m +\frac{1}{2})}\hat{a}_k^\dag\hat{a}_l\,,
\end{split}
\end{align}
and the corresponding conjugate relations, which follow from~\eqref{eq:action-U_W-on-E}.

With these relations, one can reduce all expectation values of polynomials of position and momentum operators on $\ket{\psi(S_1,S_2,M)}$ to linear combinations of terms of the form
\begin{equation}
    \braket{0|\Ug{S} \hat{\mathbf{x}}_{i_1}\cdots\hat{\mathbf{x}}_{i_n}|0}\,,
    \label{eq:gaussian-expectation}
\end{equation}
where $\Ug{S}$ is an appropriate Gaussian unitary, obtained by using~\eqref{eq:group-multiplication-gaussian} to combine all unitaries remaining after the commutations. We will now deal with the calculation of quantities of the form~\eqref{eq:gaussian-expectation}.

\subsection{Efficient computation of expectation values in standard form}\label{app:efficient-computation-ex-gaussian}
To compute the BCH decomposition~\eqref{eq:BCH} in the case of bosonic Gaussian states, it is convenient to first perform an intermediate step. Given a unitary $\Ug{S}$, we can always use the Cartan decomposition~\cite{hackl2020bosonic}
\begin{equation}
    \Ug{S}=\Ug{u^{-1}T}=\Ugd{u}\,\Ug{T}\,,
\end{equation}
with $u$ and $T$ satisfying
\begin{align}
    \Ug{u}\ket{0}=e^{\ii \theta}\ket{0}\quad\mbox{and}\quad\Omega T=T^{-1}\Omega\,,
    \label{eq:conditions-T-u}
\end{align}
where $\Omega$ was defined in~\eqref{eq:Omega}. These requirements actually fix a unique solution given\footnote{Indeed, considering that $T$ should also be an element of $\mathrm{Sp}(2N,\mathbb{R})$, \ie $T^\intercal\Omega T=\Omega$, we have that $\Omega T=T^{-1}\Omega$ implies $T=T^\intercal$. The condition $\Ug{u}\ket{0}=e^{\ii \theta}\ket{0}$ on the other hand implies $u u^\intercal=\id$, as can be seen by considering
\begin{align*}
\begin{split}
    \id&=2 \mathrm{Re}\braket{0|\hat{\mathbf{x}} \hat{\mathbf{x}}^\intercal |0}=2\mathrm{Re}\braket{0|\Ugd{u}\, \hat{\mathbf{x}} \hat{\mathbf{x}}^\intercal \,\Ug{u}|0}\\
    &=u\,\left(2\mathrm{Re}\braket{0|\hat{\mathbf{x}} \hat{\mathbf{x}}^\intercal |0}\right)\,u^{\intercal}=u u^\intercal\,.
\end{split}
\end{align*}
Using these two properties one immediately has $S^\intercal S= T^2$.} by $T=\sqrt{S^{^\intercal}\! S}$ and $u=TS^{-1}$. The phase $\theta$ can be computed as
\begin{equation}
    \theta=-\ii \braket{0|\hat{Q}(\log u)|0}=\frac{1}{4}\mathrm{tr}(\Omega \log u)\,.
\end{equation}

This decomposition means that the expectation value of interest~\eqref{eq:gaussian-expectation} can be written as
\begin{align}
\begin{split}
    \hspace{-8pt}\braket{0|\,\Ug{S} \hat{\mathbf{x}}_{i_1}\cdots\hat{\mathbf{x}}_{i_n}|0}&=e^{-\ii \theta}\braket{0|\,\Ug{T} \hat{\mathbf{x}}_{i_1}\cdots\hat{\mathbf{x}}_{i_n}|0}\\
    &=e^{-\ii \theta}\braket{0|\,e^{\hat{Q}(K)} \hat{\mathbf{x}}_{i_1}\cdots\hat{\mathbf{x}}_{i_n}|0},
\end{split}
\end{align}
where we have written $T=e^K$, with the condition~\eqref{eq:conditions-T-u} on $T$ being equivalent to $\{K,\Omega\}=0$. Considering that $K$ is also in $\mathfrak{sp}(2N)$, it must have the form 
\begin{equation}
    K=\left(\begin{array}{cc} A & B \\ B & -A \end{array}\right)
\end{equation}
with $A$ and $B$ being real symmetric $N\times N$ matrices. We therefore find
\begin{equation}
    \hat{Q}(K)= \ii \left( (K_+)_{kl} \,\hat{a}^\dag_k \,\hat{a}^\dag_l + (K_+^*)_{kl} \, \:\hat{a}_k \hat{a}_l\right)
    \label{eq:K-plus}
\end{equation}
with $K_+=\frac{1}{2}(B-\ii A)$.

We now see the purpose of the intermediate decomposition of the unitary $\Ug{S}$. This is because only for an operator of the form~\eqref{eq:K-plus}, we know how to perform the splitting~\eqref{eq:BCH} analytically, as we have~\cite{truax_baker-campbell-hausdorff_1985,hackl2020bosonic,windt_local_2020}
\begin{equation}
    e^{\hat{Q}(K)}=e^{(A_+)_{kl}\,\hat{a}^\dag_k \,\hat{a}^\dag_l} \,e^{(A_0)_{kl} \,\hat{a}^\dag_k \,\hat{a}_l + (A_0^\intercal)_{kl}\,\hat{a}_k \,\hat{a}^\dag_l} \,e^{-(A_+^*)_{kl} \,\hat{a}_k \hat{a}_l}\,,
    \label{eq:gaussian-bch}
\end{equation}
where $A_+$ is defined by the relation
\begin{align}
\begin{split}
    2 \left(\begin{array}{cc} \mathrm{Re} A_+ & \mathrm{Im} A_+ \\
    \mathrm{Im} A_+ & -\mathrm{Re} A_+ \end{array}\right)&= \tanh K = \tanh \log T \\
    &= (S^\intercal S -\id)(S^\intercal S +\id)^{-1}
\end{split}\label{eq:gaussian-A-plus}
\end{align}
and $A_0$ is calculated as
\begin{equation}
    A_0=\frac{1}{4}\log (\id-4 A_+ A_+^*)\,.
\end{equation}

As before, we see that of the three exponentials appearing in the RHS of equation~\eqref{eq:gaussian-bch} the first one acts on the lowest weight state $\bra{0}$ to its left as the identity, the second one is the exponential of operators, for which $\bra{0}$ is an eigenstate, and the third one can be commuted through the operators $\hat{\mathbf{x}}_i$ to act as the identity on the lowest weight state $\ket{0}$ to its right. To do these commutations, we use~\eqref{eq:commute-left-non-unitary}, which here takes the form
\begin{equation}
    e^{-(A_+^*)_{kl} \,\hat{a}_k \hat{a}_l} \,\hat{\mathbf{x}}_i = \mathbf{R}_{ij} \hat{\mathbf{x}}_j \, e^{-(A_+^*)_{kl} \,\hat{a}_k \hat{a}_l}\,,
\end{equation}
where $\mathbf{R}$ is the $2N\times 2N$ matrix
\begin{equation}
    \mathbf{R}=\left(\begin{array}{cc} \id-A_+^* & -\ii A_+^* \\ -\ii A_+^* & \id+A_+^* \end{array}\right)\,.
    \label{eq:R-gaussian}
\end{equation}

Combining these observations, we have the final result
\begin{equation}
    \braket{0|\,\Ug{S} \hat{\mathbf{x}}_{i_1}\cdots\hat{\mathbf{x}}_{i_n}|0}= r_0 \mathbf{R}_{i_1 j_1 }\cdots \mathbf{R}_{i_n j_n}\! \braket{0| \hat{\mathbf{x}}_{j_1}\cdots\hat{\mathbf{x}}_{j_n}|0},
\end{equation}
where $\mathbf{R}$ is given by~\eqref{eq:R-gaussian}, $A_+$ by~\eqref{eq:gaussian-A-plus} and
\begin{align}
\begin{split}
    r_0&=\exp\left(-\ii\theta+\frac{1}{4}\mathrm{tr} \log (\id-4 A_+ A_+^*)\right)\\
    &=e^{-\frac{\ii}{4}\mathrm{tr}(\Omega \log\sqrt{S^{^\intercal}\! S}S^{-1})}\det(\id-4 A_+ A_+^*)^{\frac{1}{4}}\,,
\end{split}
\end{align}
while $\braket{0| \hat{\mathbf{x}}_{j_1}\cdots\hat{\mathbf{x}}_{j_n}|0}$ can be evaluated simply with Wick's theorem.

\section{Fermionic Gaussian states}\label{app:ex-fermionic-gaussian}
We now consider the case of fermionic Gaussian states. This example complements the previous one of bosonic Gaussian states, giving the reader an indication of how to apply our constructions to even more general settings, \ie the ones which include fermions. As before, we follow the structure of the main body of the paper illustrating the construction step-by-step.

\subsection{Group-theoretic coherent states}\label{app:group-theoretic-coherent-states-ex-fermionic-gaussian}
We consider a system of $N$ fermionic modes, characterized by the annihilation and creation operators $\hat{c}_1,\dots,\hat{c}_N,\hat{c}^\dag_1,\dots,\hat{c}^\dag_N$. It is useful to also consider the Hermitian operators $\hat{\gamma}_k=\frac{1}{\sqrt{2}}(\hat{c}^\dag_k+\hat{c}_k)$ and $\hat{\bar{\gamma}}_k=\frac{\ii}{\sqrt{2}}(\hat{c}^\dag_k-\hat{c}_k)$, which are typically referred to as Majorana operators. They play a role analogous to the one of position and momentum operators in the bosonic case. They satisfy the anti-commutation relations
\begin{equation}
    \{\hat{\gamma}_k,\hat{\gamma}_l\}=\{\hat{\bar{\gamma}}_k,\hat{\bar{\gamma}}_l\}=\delta_{kl}, \quad \{\hat{\gamma}_k,\hat{\bar{\gamma}}_l\}=0\,.
    \label{eq:fermionic-gaussian-commutators}
\end{equation}

Gaussian unitaries are defined as operators of the form $\mathcal{U}=e^{\hat{Q}}$, where $\hat{Q}$ is any anti-Hermitian homogeneous order 2 polynomial in the operators $\hat{\gamma}_k,\hat{\bar{\gamma}}_k$. More precisely, if we group all the Majorana operators into a single $2N$-dimensional vector $\hat{\mathbf{x}}=(\hat{\gamma}_1,\dots,\hat{\gamma}_N,\hat{\bar{\gamma}}_1,\dots,\hat{\bar{\gamma}}_N)^\intercal$, $\hat{Q}$ can be put in the form
\begin{equation}
    \hat{Q}=\frac{1}{2} \: \hat{\mathbf{x}}^\intercal K \hat{\mathbf{x}}\,,
\end{equation}
where $K$ is any $2N\times 2N$ real anti-symmetric matrix. In principle $K$ could be any anti-Hermitian matrix. However, using the anti-commutation relations~\eqref{eq:fermionic-gaussian-commutators} one can show that the symmetric part of $K$ only contributes an imaginary c-number to $\hat{Q}$, therefore only a global phase to $\mathcal{U}$, in which we are not interested. So we can assume $K$ to be anti-symmetric and real.

Gaussian states are defined as the states obtained by acting with any Gaussian unitary on the Fock vacuum $\ket{0}$. Thus, Gaussian states are all of the form $e^{\hat{Q}}\ket{0}$ for any allowed $\hat{Q}$. Here, the vacuum is defined as the state annihilated by all annihilation operators, \ie $\hat{c}_k\ket{0}=0$, $\forall k$.

Fermionic Gaussian states defined in this way fit into the group-theoretic coherent states formalism described in section~\ref{sec:group-theoretic-coherent-states-global}. This is because the fermionic Gaussian operators $\mathcal{U}$ that we have defined give a unitary representation of the Lie group of real orthogonal matrices
\begin{align}
    \mathrm{O}(2N,\mathbb{R})=\{\G\in\mathrm{GL}(2N,\mathbb{R})\st \G^\intercal \G=\id\}\,.
\end{align}
Similarly, the set of anti-Hermitian operators $\hat{Q}$ give a representation of the Lie algebra of anti-symmetric matrices
\begin{align}
    \mathfrak{so}(2N,\mathbb{R})&=\{K\in\mathfrak{gl}(2N,\mathbb{R})\st K+K^\intercal=0\}\,.
\end{align}
Indeed, for each matrix $K\in\mathfrak{so}(2N,\mathbb{R})$, one can construct the corresponding Hilbert space operator
\begin{equation}
    \hat{Q}(K)=\frac{1}{2} \:\hat{\mathbf{x}}^\intercal K \hat{\mathbf{x}}\,.
    \label{eq:fermionic-gaussian-Q}
\end{equation}
Similarly, for any matrix $\G\in\mathrm{O}(2N,\mathbb{R})$ that can be written as $\G=e^K$ for some $K\in\mathfrak{so}(2N,\mathbb{R})$, one can define the corresponding unitary
\begin{equation}
    \mathcal{U}(\G)=\mathcal{U}(e^{K})=e^{\hat{Q}(K)}\,.
    \label{eq:fermionic-gaussian-unitary}
\end{equation}
The operators $\mathcal{U}(\G)$ constitute a group representation, in the sense that one can show that
\begin{equation}
    \mathcal{U}(\G)\,\mathcal{U}(\tilde{\G})=\mathcal{U}(\G\tilde{\G})\,.
    \label{eq:group-multiplication-fermionic-gaussian}
\end{equation}

As in section~\ref{sec:group-theoretic-coherent-states-global}, the algebra operators $\hat{Q}$ defined in~\eqref{eq:fermionic-gaussian-Q} can be expanded on a basis $\hat{\X}_i$. In this case, $\hat{Q}$ can be expanded as
\begin{align}
\begin{split}
    \hat{Q}&=A^{kl}\frac{1}{2}(\hat{\gamma}_k\hat{\gamma}_l+\hat{\bar{\gamma}}_k\hat{\bar{\gamma}}_l)+B^{kl}\frac{1}{2}(\hat{\gamma}_k\hat{\gamma}_l-\hat{\bar{\gamma}}_k\hat{\bar{\gamma}}_l)\\
    &\hspace{10pt}+ C^{kl}\frac{1}{2}(\hat{\gamma}_k\hat{\bar{\gamma}}_l+\hat{\bar{\gamma}}_k\hat{\gamma}_l)+D^{kl}\frac{1}{2}(\hat{\gamma}_k\hat{\bar{\gamma}}_l-\hat{\bar{\gamma}}_k\hat{\gamma}_l)\,,
\end{split}
\end{align}
for real anti-symmetric $A^{kl}$, $B^{kl}$, $C^{kl}$ and real symmetric $D^{kl}$. Thus, all $\hat{Q}$ are real linear combinations of the operators
\begin{subequations}
\begin{align}
    \frac{1}{2}(\hat{\gamma}_k\hat{\bar{\gamma}}_k-\hat{\bar{\gamma}}_k\hat{\gamma}_k)&=\ii(\hat{c}_k^\dag \hat{c}_k-\frac{1}{2})\label{eq:Z-basis-fermionic-gaussian-first}\\
    \frac{1}{2}(\hat{\gamma}_k\hat{\bar{\gamma}}_l-\hat{\bar{\gamma}}_k\hat{\gamma}_l)&=\frac{\ii}{2}(\hat{c}_k^\dag \hat{c}_l-\hat{c}_k \hat{c}_l^\dag ),\quad k< l\\
    \frac{1}{2}(\hat{\gamma}_k\hat{\gamma}_l+\hat{\bar{\gamma}}_k\hat{\bar{\gamma}}_l)&=\frac{1}{2}(\hat{c}_k \hat{c}_l^\dag+\hat{c}_k^\dag \hat{c}_l ),\quad k< l\\
    \frac{1}{2}(\hat{\gamma}_k\hat{\gamma}_l-\hat{\bar{\gamma}}_k\hat{\bar{\gamma}}_l)&= \frac{1}{2}(\hat{c}_k^\dag\hat{c}_l^\dag+\hat{c}_k\hat{c}_l),\quad k< l\\
    \frac{1}{2}(\hat{\gamma}_k\hat{\bar{\gamma}}_l+\hat{\bar{\gamma}}_k\hat{\gamma}_l)&= \frac{\ii}{2}(\hat{c}_k^\dag\hat{c}_l^\dag-\hat{c}_k\hat{c}_l),\quad k< l \label{eq:Z-basis-fermionic-gaussian-last}
\end{align}
\end{subequations}
which play the role of the operators $\hat{\X}_i$.

These can in turn be decomposed into combinations of Cartan subalgebra operators $\hat{H}_\I$ and root space operators $\hat{E}_\al$. More specifically, we can choose Cartan operators
\begin{equation}
    \hat{H}_k=\ii(\hat{c}_k^\dag \hat{c}_k -\tfrac{1}{2})\,,
    \label{eq:cartan-operators-fermionic-gaussian}
\end{equation}
which leads to the root space operators
\begin{subequations}
\begin{align}
    \hat{E}_{+\al^{(k,l)}}&=\hat{c}^\dag_k\hat{c}^\dag_l,&\hat{E}_{-\al^{(k,l)}}&=\hat{c}_k\hat{c}_l,& k\leq l\\
    \hat{E}_{+\tilde{\al}^{(k,l)}}&=\hat{c}^\dag_k\hat{c}_l,&\hat{E}_{-\tilde{\al}^{(k,l)}}&=\hat{c}_k\hat{c}^\dag_l,&k<l
\end{align}
\end{subequations}
corresponding to the root vectors $\al^{\!(k,l)}_{\,\I}=(\delta_{ak}+\delta_{al})$ and $\tilde{\al}^{(k,l)}_{\,\I}=(\delta_{\I k}-\delta_{\I l})$. We see by inspection that all algebra operators $\hat{\X}_i$ as defined in equations~\eqref{eq:Z-basis-fermionic-gaussian-first} to~\eqref{eq:Z-basis-fermionic-gaussian-last} are complex linear combinations of these objects. The Fock vacuum $\ket{0}$ is the corresponding lowest weight state. Indeed, it is an eigenstate with eigenvalue $-\frac{\ii}{2}$ of all Cartan subalgebra operators $\hat{H}_k$ and it is annihilated by all negative root space operators $\hat{E}_{-\al^{(k,l)}}\ket{0}=\hat{E}_{-\tilde{\al}^{(k,l)}}\ket{0}=0$.

We conclude that fermionic Gaussian states fulfil all the criteria to be identified as the group-theoretic coherent states for the group $\mathcal{G}=\mathrm{O}(2N,\mathbb{R})$, given its unitary representation in terms of fermionic operators described above.

\subsection{Generalized family of states}\label{app:generalized-family-ex-fermionic-gaussian}
We now construct generalized fermionic Gaussian states following our definition in~\ref{sec:generalized-family}.  Based on~\eqref{eq:cartan-operators-fermionic-gaussian}, we choose our Cartan subalgebra operators as
\begin{equation}
    \hat{H}_k=\ii(\hat{c}_k^\dag \hat{c}_k -\frac{1}{2})\,.
\end{equation} 
This leads to the non-Gaussian unitaries of the form
\begin{equation}
    \Ung{M}=\exp \left(-\frac{i}{2} M^{kl} (\hat{c}_k^\dag \hat{c}_k -\frac{1}{2})(\hat{c}_l^\dag \hat{c}_l -\frac{1}{2}) \right)
\end{equation}
for any $N\times N$ real symmetric matrix $M$. The \emph{generalized bosonic Gaussian states} are then defined as
\begin{equation}
    \ket{\psi(\G_1,\G_2,M)}=\Ug{\G_1}\,\Ung{M}\,\Ug{\G_2}\ket{0}\,,
\end{equation}
where $\Ug{\G}$ are the Gaussian unitaries defined in~\eqref{eq:fermionic-gaussian-unitary}. We recognize that these states constitute one of the classes of non-Gaussian states previously introduced in~\cite{shi2018variational}, which is not surprising as this construction heavily inspired us to define generalized group-theoretic coherent states in the prescribed way.

In this setting, the observables of interest will be polynomials in the operators $\hat{\gamma}_k$ and $\hat{\bar{\gamma}}_k$, or equivalently in $\hat{c}_k^\dag$ and $\hat{c}_k$. As before, in order to compute expectation values of such observables, we need to commute them with unitaries of the types $\Ug{\G}$ and $\Ung{M}$. This can be achieved thanks to the relations 
\begin{equation}
    \Ugd{\G}\,\hat{\mathbf{x}}\,\Ug{\G}=\G\hat{\mathbf{x}}\,,
\end{equation}
which can be derived from~\eqref{eq:fermionic-gaussian-commutators}, and
\begin{align}
\begin{split}
    \hspace{-7pt}\Ungd{M}\hat{c}_k\hat{c}_l \Ung{M}&=e^{-\frac{\ii}{2}(M^{kk}+M^{kl}+M^{lk}+M^{ll})}\\
    &\times e^{-\ii(M^{km}+M^{lm})(\hat{c}_m^\dag \hat{c}_m +\frac{1}{2})}\hat{c}_k\hat{c}_l\,,
\end{split}\\
\begin{split}
    \hspace{-7pt}\Ungd{M}\hat{c}_k^\dag\hat{c}_l \Ung{M}&=e^{-\frac{\ii}{2}(M^{kk}-M^{kl}-M^{lk}+M^{ll})}\\
    &\hspace{5pt}\times e^{\ii(M^{km}-M^{lm})(\hat{c}_m^\dag \hat{c}_m +\frac{1}{2})}\hat{c}_k^\dag\hat{c}_l\,,
\end{split}
\end{align}
and the corresponding conjugate relations, which follow from~\eqref{eq:action-U_W-on-E}.

With these relations, one can reduce all expectation values of polynomials of Majorana operators on $\ket{\psi(\G_1,\G_2,M)}$ to linear combinations of terms of the form
\begin{equation}
    \braket{0|\Ug{\G} \hat{\mathbf{x}}_{i_1}\cdots\hat{\mathbf{x}}_{i_n}|0}\,,
    \label{eq:fermionic-gaussian-expectation}
\end{equation}
where $\Ug{\G}$ is an appropriate Gaussian unitary, obtained by using~\eqref{eq:group-multiplication-fermionic-gaussian} to combine all unitaries remaining after the commutations. We will now deal with the calculation of quantities of the form~\eqref{eq:fermionic-gaussian-expectation}.

\subsection{Efficient computation of expectation values in standard form}\label{app:efficient-computation-ex-fermionic-gaussian}
To compute the BCH decomposition~\eqref{eq:BCH} in the case of fermionic Gaussian states, it is convenient to first perform an intermediate step. Given a unitary $\Ug{\G}$, we can always use the Cartan decomposition~\cite{hackl2020bosonic}
\begin{equation}
    \Ug{\G}=\Ug{u^{-1}T}=\Ugd{u}\,\Ug{T}\,,
\end{equation}
with $u$ and $T$ satisfying
\begin{align}
    \Ug{u}\ket{0}=e^{\ii \theta}\ket{0}\quad\mbox{and}\quad\Omega T=T^{-1}\Omega\,,
    \label{eq:conditions-T-u-fermions}
\end{align}
where $\Omega$ was defined in~\eqref{eq:Omega}. These requirements actually fix a unique solution given\footnote{Indeed, considering that $T$ should also be an element of $\mathrm{O}(2N,\mathbb{R})$, \ie $T^{^\intercal}\! =T^{-1}$, we have that $\Omega T=T^{-1}\Omega$ implies $\Omega T=T^{^\intercal}\! \Omega$. The condition $\Ug{u}\ket{0}=e^{\ii \theta}\ket{0}$ on the other hand implies $u \Omega u^\intercal=\Omega$, as can be seen by considering
\begin{align*}
\begin{split}
    \Omega&=2 \mathrm{Im}\braket{0|\hat{\mathbf{x}} \hat{\mathbf{x}}^\intercal |0}=2\mathrm{Im}\braket{0|\Ugd{u}\, \hat{\mathbf{x}} \hat{\mathbf{x}}^\intercal \,\Ug{u}|0}\\
    &=u\,\left(2\mathrm{Im}\braket{0|\hat{\mathbf{x}} \hat{\mathbf{x}}^\intercal |0}\right)\,u^{\intercal}=u \Omega  u^\intercal\,.
\end{split}
\end{align*}
Using these two properties one immediately has $-\Omega \G^\intercal \Omega \G= T^2$.} by $T=\sqrt{-\Omega \G^{^\intercal}\!\Omega \G}$ and $u=T\G^{-1}$. The phase $\theta$ can be computed as
\begin{equation}
    \theta=-\ii \braket{0|\hat{Q}(\log u)|0}=-\frac{1}{4}\mathrm{tr}(\Omega \log u)\,.
\end{equation}

This decomposition means that the expectation value of interest~\eqref{eq:fermionic-gaussian-expectation} can be written as
\begin{align}
\begin{split}
    \hspace{-8pt}\braket{0|\,\Ug{\G} \hat{\mathbf{x}}_{i_1}\cdots\hat{\mathbf{x}}_{i_n}|0}&=e^{-\ii \theta}\braket{0|\,\Ug{T} \hat{\mathbf{x}}_{i_1}\cdots\hat{\mathbf{x}}_{i_n}|0}\\
    &=e^{-\ii \theta}\braket{0|\,e^{\hat{Q}(K)} \hat{\mathbf{x}}_{i_1}\cdots\hat{\mathbf{x}}_{i_n}|0},
\end{split}
\end{align}
where we have written $T=e^K$, with the condition~\eqref{eq:conditions-T-u-fermions} on $T$ being equivalent to $\{K,\Omega\}=0$. Considering that $K$ is also in $\mathfrak{so}(2N)$, it must have the form 
\begin{equation}
    K=\left(\begin{array}{cc} A & B \\ B & -A \end{array}\right)
\end{equation}
with $A$ and $B$ being real anti-symmetric $N\times N$ matrices. We therefore find
\begin{equation}
    \hat{Q}(K)= \left( (K_+)_{kl} \,\hat{c}^\dag_k \,\hat{c}^\dag_l + (K_+^*)_{kl} \, \:\hat{c}_k \hat{c}_l\right)
    \label{eq:K-plus-fermionic}
\end{equation}
with $K_+=\frac{1}{2}(A+\ii B)$.

We now see the purpose of the intermediate decomposition of the unitary $\Ug{\G}$. This is because only for an operator of the form~\eqref{eq:K-plus-fermionic}, we know how to perform the splitting~\eqref{eq:BCH} analytically, as we have~\cite{truax_baker-campbell-hausdorff_1985,hackl2020bosonic,windt_local_2020}
\begin{equation}
    e^{\hat{Q}(K)}=e^{(A_+)_{kl}\,\hat{c}^\dag_k \,\hat{c}^\dag_l} \,e^{(A_0)_{kl} \,\hat{c}^\dag_k \,\hat{c}_l - (A_0^\intercal)_{kl}\,\hat{c}_k \,\hat{c}^\dag_l} \,e^{(A_+^*)_{kl} \,\hat{c}_k \hat{c}_l}\,,
    \label{eq:fermionic-gaussian-bch}
\end{equation}
where $A_+$ is defined by the relation
\begin{align}
\begin{split}
    2 \left(\begin{array}{cc} \mathrm{Re} A_+ & \mathrm{Im} A_+ \\
    \mathrm{Im} A_+ & -\mathrm{Re} A_+ \end{array}\right)&= \tanh K = \tanh \log T \\
    &= (\Omega \G^\intercal \Omega \G +\id)(\Omega \G^\intercal \Omega \G -\id)^{-1}
\end{split}\label{eq:fermionic-gaussian-A-plus}
\end{align}
and $A_0$ is calculated as
\begin{equation}
    A_0=\frac{1}{4}\log (\id-4 A_+ A_+^*)\,.
\end{equation}

As before, we see that of the three exponentials appearing in the RHS of equation~\eqref{eq:fermionic-gaussian-bch} the first one acts on the lowest weight state $\bra{0}$ to its left as the identity, the second one is the exponential of operators, for which $\bra{0}$ is an eigenstate, and the third one can be commuted through the operators $\hat{\mathbf{x}}_i$ to act as the identity on the lowest weight state $\ket{0}$ to its right. To do these commutations, we use~\eqref{eq:commute-left-non-unitary}, which here takes the form
\begin{equation}
    e^{(A_+^*)_{kl} \,\hat{c}_k \hat{c}_l} \,\hat{\mathbf{x}}_i = \mathbf{R}_{ij} \hat{\mathbf{x}}_j \, e^{(A_+^*)_{kl} \,\hat{c}_k \hat{c}_l}\,,
\end{equation}
where $\mathbf{R}$ is the $2N\times 2N$ matrix
\begin{equation}
    \mathbf{R}=\left(\begin{array}{cc} \id-A_+^* & -\ii A_+^* \\ -\ii A_+^* & \id+A_+^* \end{array}\right)\,.
    \label{eq:R-fermionic-gaussian}
\end{equation}

Combining these observations, we have the final result
\begin{equation}
    \braket{0|\,\Ug{\G} \hat{\mathbf{x}}_{i_1}\cdots\hat{\mathbf{x}}_{i_n}|0}= r_0 \mathbf{R}_{i_1 j_1 }\cdots \mathbf{R}_{i_n j_n}\! \braket{0| \hat{\mathbf{x}}_{j_1}\cdots\hat{\mathbf{x}}_{j_n}|0},
\end{equation}
where $\mathbf{R}$ is given by~\eqref{eq:R-fermionic-gaussian}, $A_+$ by~\eqref{eq:fermionic-gaussian-A-plus} and
\begin{align}
\begin{split}
    r_0&=\exp\left(-\ii\theta-\frac{1}{4}\mathrm{tr} \log (\id-4 A_+ A_+^*)\right)\\
    &=e^{\frac{\ii}{4}\mathrm{tr}(\Omega \log\sqrt{-\Omega \G^{^\intercal}\!\Omega \G}\G^{-1})}\det(\id-4 A_+ A_+^*)^{-\frac{1}{4}}\,,
\end{split}
\end{align}
while $\braket{0| \hat{\mathbf{x}}_{j_1}\cdots\hat{\mathbf{x}}_{j_n}|0}$ can be evaluated simply with Wick's theorem.

\section{Variational methods}\label{app:variational-methods}
The main application of a family of states $\ket{\psi(x)}$ such as the one defined in~\eqref{eq:definition-generalized-coherent-states} (where we indicate with $x$ collectively all the parameters defining the state) is to use it as the \emph{ansatz} for a variational calculation. In this appendix we show that all the relevant quantities one needs to compute for such application can be brought to linear combinations of terms of the form
\begin{equation}
    \braket{\mu|\,\Ug{\M}\,\hat{\X}_{i_1}\cdots\hat{\X}_{i_n}|\mu} \,.
    \label{eq:expectation-U-Xi-app}
\end{equation}
To do this we use the result of section~\ref{sec:standard-form} that the adjoint action of $\Ung{M}$ on any polynomial of operators $\hat{\X}_i$ gives rise to a linear combination of products of group operations and algebra operators.

Given a Hamiltonian $\hat{H}$ defined on $\mathcal{H}$, an \emph{ansatz} $\ket{\psi(x)}$ may be used both to approximate the ground state of $\hat{H}$ and to simulate the real time dynamics of the system. This can be done according to different variational principles, as discussed in~\cite{hackl2020geometry} and illustrated for Gaussian states in \cite{kraus2010generalized,shi2018variational,guaita2019gaussian}. To do so it is necessary to be able to compute the following quantities:
\begin{align}
    \braket{\psi(x)|\hat{H}|\psi(x)},\,
    \braket{V_\mu(x)|\hat{H}|\psi(x)},\,
    \braket{V_\mu(x)|V_\nu(x)}\label{eq:variational-quantities},
\end{align}
where $\ket{V_\mu(x_0)}=\frac{\partial}{\partial x^\mu}\ket{\psi(x)}|_{x=x_0}$ is a so-called \emph{tangent vector} of the variational manifold.

Here, we have assumed that the group $\mathcal{G}$ and its representation have been chosen so that $\hat{H}$ can be expressed as a polynomial in the operators $\hat{\X}_i$.
For what concerns the computation of the tangent vectors, it can be shown (see~\cite{hackl2020geometry}) that the derivatives of $\Ug{\M}$ with respect to a suitable parametrization of the group can be written as linear combinations of terms of the form $\Ug{\M}\,\hat{\X}_i$. Similarly, it holds that
\begin{equation}
    \frac{\partial}{\partial \W^{\I\J}} \Ung{\W}=\frac{\ii}{2}\,\Ung{\W} \hat{H}_\I \hat{H}_\J\,.
\end{equation}
Consequently, we have that for generalized group-theoretic coherent states tangent vectors have the form
\begin{align}
\begin{split}
    \ket{V_\mu(x)}=&C_1^{i}\; \Ug{\M_1}\hat{\X}_i\Ung{\W}\Ug{\M_2}\ket{\mu}\\
    &+C_2^{i}\;\Ug{\M_1}\Ung{\W}\Ug{\M_2}\hat{\X}_i\ket{\mu}\\
    &+C_3^{\I\J}\;  \Ug{\M_1}\Ung{\W}\hat{H}_\I\hat{H}_\J\Ug{\M_2}\ket{\mu}\,.
\end{split}
\end{align}

With this in mind, one sees immediately that the quantities~\eqref{eq:variational-quantities} are made up of terms where one has to evaluate repeatedly the adjoint action of $\Ug{\M}$ or $\Ung{\W}$ on products of operators $\hat{\X}_i$ and then compute the expectation value of the result on $\ket{\mu}$. Using the results~\eqref{eq:action-U_W-on-E} and \eqref{eq:action-U_W-on-H} these give rise to linear combinations of further products of operators $\hat{\X}_i$ and potentially of group transformations $\Ug{\M}$. Using then~\eqref{eq:commute-left} as explained in Section~\ref{sec:standard-form} to commute all the group transformations to the left, they can thus be all brought to linear combinations of terms of the form~\eqref{eq:expectation-U-Xi-app}.
\newline

\section{Computation of \texorpdfstring{$\mathrm{Ad}(\M)_i^j$}{Ad(M)}}
\label{app:adjoint-rep}
The object $\mathrm{Ad}(\M)_i^j$ plays an important role for our calculation of arbitrary expectation values. It is known in mathematics as the adjoint action of the group onto its algebra. If we represent the group element $g$ and the algebra element $Z_i$ by matrices (in any representation), we have
\begin{align}
    g^{-1} Z_i g=\mathrm{Ad}(g)_i^j \,Z_j\,.
    \label{eq:matrix-ajoint}
\end{align}
In other words the product of matrices $g^{-1} Z_i g$ can be reexpressed as a linear combination of algebra matrices with coefficients $\mathrm{Ad}(g)_i^j$. These coefficients are independent of the chosen representation. Consequently, the calculation is done very efficiently if we choose a representation with a low dimension, such as the fundamental representation.

To extract the  coefficients $(\mathrm{Ad}_g)_i{}^j$, we can use the fact that $\kappa_{ij}=\mathrm{Tr}(Z_iZ_j^\dagger)$ is a positive-definite inner product on the space of matrices in whatever basis we represent them. We then represent the group element $g$ and the Lie algebra elements $Z_i$ as matrices, multiply~\eqref{eq:matrix-ajoint} by $Z_k^\dag$, take the trace and have
\begin{equation}
    \mathrm{Ad}(g)_i^j \kappa_{jk}=\mathrm{Tr}(g^{-1}Z_igZ_k^\dagger)\,.
\end{equation}
multilpying by the inverse of $\kappa$ we finally have
\begin{align}
    \mathrm{Ad}(g)_i^j=\mathrm{Tr}(g^{-1}Z_igZ_k^\dagger)(\kappa^{-1})^{kj}\,.
\end{align}
If we do this numerically, we only need to compute $\kappa^{-1}$ once and find that the matrix $\mathrm{Ad}(g)_i^j$ can be efficiently computed for any group element $g$.

Alternatively, if the group element $g$ can be represented as $e^{K^i Z_i}$, then from a well known Baker-Campbell-Hausdorff relation we have
\begin{align}
\begin{split}
    g^{-1} Z_i g&= e^{-K^j Z_j} Z_i e^{K^k Z_k}\\
    &=Z_i+ [K^j Z_j,Z_i] +\frac{1}{2!}[K^j Z_j,[K^k Z_k,Z_i]]+\cdots \\
    &=Z_i + K^j c_{ji}^l Z_l +\frac{1}{2!} K^k c_{ki}^l K^j c_{jl}^m Z_m+\cdots\\
    &=(e^{\mathrm{ad}(K)})^j_i Z_j \,.
\end{split}
\end{align}
where $\mathrm{ad}(K)^j_i=K^k c_{ki}^j$. This shows that the result only depends on the algebra commutation relations and not on the specific representation. It is also useful in the case in which $K^i$ are complex coefficients, as in the derivation of~\eqref{eq:commute-left-non-unitary}.

\bibliography{main}

\end{document}